\documentclass[prb,twocolumn,epsf,floats,aps,eqsecnum,showpacs]{revtex4-1}

\usepackage{amsmath}
\usepackage{amssymb}
\usepackage{mathtools}
\usepackage{wrapfig}
\usepackage{enumerate}
\usepackage{setspace}
\usepackage[hypertex]{hyperref}
\usepackage[caption=false]{subfig}
\usepackage{ytableau}
\usepackage{bbm}
\usepackage{graphics}
\usepackage{graphicx}% Include Fig. files

\newcommand{\one}{\text{\small 1}\!\!1}

\newcommand{\be}{\begin{equation}}
\newcommand{\ee}{\end{equation}}
\newcommand{\bea}{\begin{eqnarray}}
\newcommand{\eea}{\end{eqnarray}}

\newcommand{\mfg}{\mathfrak g}
\newcommand{\mfh}{\mathfrak h}
\newcommand{\mfn}{\mathfrak n}
\newcommand{\mfp}{\mathfrak p}
\newcommand{\mfk}{\mathfrak k}

\begin{document}

\title{Classification and symmetry properties of scaling dimensions at Anderson transitions}

\author{I.~A.~Gruzberg}
\affiliation{The James Franck Institute and the Department of Physics, The University of Chicago, Chicago, Illinois 60637, USA}
\author{A.~D.~Mirlin}
\affiliation{Institut f\"ur Nanotechnologie, Karlsruhe Institute of Technology, 76021 Karlsruhe, Germany}
\affiliation{Institut f\"ur Theorie der kondensierten Materie and DFG Center for Functional Nanostructures, Karlsruhe Institute of Technology, 76128 Karlsruhe, Germany}
\affiliation{Petersburg Nuclear Physics Institute, 188300 St.~Petersburg, Russia}
\author{M.~R.~Zirnbauer}
\affiliation{Institut f\"ur Theoretische Physik, Universit\"at zu K\"oln, Z\"ulpicher Stra{\ss}e 77, 50937 K\"oln, Germany}

\date{October 24, 2012}

\begin{abstract}
We develop a classification of composite operators without gradients at Anderson-transition critical points in disordered systems. These operators represent correlation functions of the local density of states (or of wave-function amplitudes). Our classification is motivated by the Iwasawa decomposition for the field of the pertinent supersymmetric $\sigma$-model: the scaling operators are represented by ``plane waves'' in terms of the corresponding radial coordinates. We also present an alternative construction of scaling operators by using the notion of highest-weight vector. We further argue that a certain Weyl-group invariance associated with the $\sigma$-model manifold leads to numerous exact symmetry relations between the scaling dimensions of the composite operators. These symmetry relations generalize those derived earlier for the multifractal spectrum of the leading operators.
\end{abstract}


\maketitle

\section{Introduction}\label{s1}

The phenomenon of Anderson localization of a quantum particle or a classical wave in a random environment is one of the central discoveries made by condensed matter physics in the second half of the last century. \cite{Anderson58} Although more than fifty years have passed since Anderson's original paper, Anderson localization remains a vibrant research field. \cite{AL50} One of its central research directions is the physics of Anderson transitions, \cite{evers08} including metal-insulator transitions and transitions of quantum-Hall type (i.e.\ between different phases of topological insulators). While such transitions are conventionally observed in electronic conductor and semiconductor structures, there is also a considerable number of other experimental realizations actively studied in recent and current works. These include localization of light, \cite{wiersma97} cold atoms, \cite{BEC-localization} ultrasound, \cite{faez09} and optically driven atomic systems. \cite{lemarie10} On the theory side, the field received a strong boost through the discovery of unconventional symmetry classes and the development of a complete symmetry classification of disordered systems. \cite{altland97, zirnbauer96, evers08, heinzner05} The unconventional classes emerge due to additional particle-hole and/or chiral symmetries that are, in particular, characteristic for models of disordered superconductors and disordered Dirac fermions (e.g.\ in graphene). In total one has 10 symmetry classes, including three standard (Wigner-Dyson) classes, three chiral, and four Bogoliubov-de Gennes (``superconducting'') classes. This multitude is further supplemented by the possibility for the underlying field theories to have a non-trivial topology ($\theta$ and Wess-Zumino terms), leading to a rich ``zoo'' of Anderson-transition critical points. The recent advent of graphene \cite{graphene-revmodphys} and of topological insulators and superconductors \cite{topins} reinforced the experimental relevance of these theoretical concepts.

In analogy with more conventional second-order phase transitions, Anderson transitions fall into different universality classes according to the spatial dimension, symmetry, and topology. In each of the universality classes, the behavior of physical observables near the transition is characterized by critical exponents determined by the scaling dimensions of the corresponding operators.

A remarkable property of Anderson transitions is that the critical wave functions are multifractal due to their strong fluctuations. Specifically, the wave-function moments show anomalous multifractal scaling with respect to the system size $L$,
\begin{equation}\label{e1.1}
 L^d \langle |\psi({\bf r})|^{2q} \rangle \propto L^{-\tau_q}, \qquad \tau_q = d(q-1) +  \Delta_q,
\end{equation}
where $d$ is the spatial dimension, $\langle \ldots \rangle$ denotes the operation of disorder averaging and $\Delta_q$ are anomalous multifractal exponents that distinguish the critical point from a simple metallic phase, where $\Delta_q \equiv 0$. Closely related is the scaling of moments of the local density of states (LDOS) $\nu(r)$,
\begin{equation}\label{e1.2}
 \langle \nu^q \rangle \propto L^{-x_q} , \qquad x_q = \Delta_q + qx_\nu,
\end{equation}
where $x_\nu \equiv x_1$ controls the scaling of the average LDOS, $\langle \nu \rangle \propto L^{-x_\nu}$. Multifractality implies the presence of infinitely many relevant (in the renormalization-group (RG) sense) operators at the Anderson-transition critical point. First steps towards the experimental determination of multifractal spectra have been made recently. \cite{faez09, lemarie10, richardella10}

Let us emphasize that when we speak about a $q$-th moment, we neither require that $q$ is an integer nor that it is positive. Throughout the paper, the term ``moment'' is understood in this broad sense.

In Refs.~[\onlinecite{mirlin94},\onlinecite{fyodorov04}] a symmetry relation for the LDOS distribution function (and thus, for the LDOS moments) in the Wigner-Dyson symmetry classes was derived:
\begin{equation}\label{e1.4}
 {\cal P}(\nu) = \nu^{-q_*-2}{\cal P}(\nu^{-1}), \qquad \langle \nu^q \rangle = \langle \nu^{q_*-q} \rangle,
\end{equation}
with $q_*=1$. Equation (\ref{e1.4}) is obtained in the framework of the non-linear $\sigma$-model and is fully general otherwise, i.e., it is equally applicable to metallic, localized, and critical systems. An important consequence of Eq.~(\ref{e1.4}) is an exact symmetry relation for Anderson-transition multifractal exponents \cite{mirlin06}
\begin{equation}\label{e1.3}
 x_q = x_{q_*-q} .
\end{equation}
While $\sigma$-models in general are approximations to particular microscopic systems, Eq.~(\ref{e1.3}) is exact in view of the universality of critical behavior.

In a recent paper, \cite{gruzberg11} the three of us and A.~W.~W.~Ludwig uncovered the group-theoretical origin of the symmetry relations (\ref{e1.4}), (\ref{e1.3}). Specifically, we showed that these relations are manifestations of a Weyl symmetry group acting on the $\sigma$-model manifold. This approach was further used to generalize these relations to the unconventional (Bogoliubov-de Gennes) classes CI and C, with $q_*=2$ and $q_*=3$, respectively.

The operators representing the averaged LDOS moments (\ref{e1.2}) by no means exhaust the composite operators characterizing LDOS (or wave-function) correlations in a disordered system. They are distinguished in that they are the dominant (or most relevant) operators for each $q$, but they only represent ``the tip of the iceberg'' of a much larger family of gradientless composite operators. Often, the subleading operators are also very important physically. An obvious example is the two-point correlation function
\begin{eqnarray}\label{e1.5}
 K_{\alpha\beta}({\bf r}_1,{\bf r}_2) &=& |\psi_\alpha^2({\bf r}_1) \psi_\beta^2({\bf r}_2)| \nonumber \\ &-& \psi_\alpha({\bf r}_1) \psi_\beta({\bf r}_2)\psi_\alpha^*({\bf r}_2)\psi_\beta^*({\bf r}_1),
\end{eqnarray}
which enters in the Hartree-Fock matrix element of a two-body interaction,
\begin{equation}\label{e1.6}
 M_{\alpha\beta} = \int dr_1 dr_2 K_{\alpha\beta}({\bf r}_1,{\bf r}_2) U({\bf r}_1 - {\bf r}_2).
\end{equation}
Questions about the scaling of the disorder-averaged function $K_{\alpha \beta} ({\bf r}_1,{\bf r}_2)$, its moments, and the correlations of such objects, arise naturally when one studies, e.g., the interaction-induced dephasing at the Anderson-transition critical point. \cite{Lee96,Wang00,burmistrov11}

The goals and the results of this paper are threefold:

\begin{enumerate}
\item We develop a systematic and complete classification of gradientless composite operators in the supersymmetric non-linear $\sigma$-models of Anderson localization. Our approach here differs from that of H\"of and Wegner \cite{Hoef86} and Wegner \cite{Wegner1987a, Wegner1987b} in two respects. Firstly, we work directly with the supersymmetric (SUSY) theories rather than with their compact replica versions as in Refs.~[\onlinecite{Hoef86, Wegner1987a, Wegner1987b}]. Secondly, we employ (a superization of) the Iwasawa decomposition and the Harish-Chandra isomorphism, which allow us to explicitly construct ``radial plane waves'' that are eigenfunctions of the Laplace-Casimir operators of the $\sigma$-model symmetry group, for arbitrary (also non-integer, negative, and even complex) values of a set of parameters $q_i$ [generalizing the order $q$ of the moment in Eq.~(\ref{e1.1}), (\ref{e1.2})]. We also develop a more basic construction of scaling operators as highest-weight vectors (and explain the link with
the Iwasawa-decomposition formalism).
\item We establish a connection between these composite operators and the physical observables of LDOS and wave-function correlators, as well as with some transport observables.
\item Furthermore, the Iwasawa-decomposition formalism allows us to exploit a certain Weyl-group invariance and deduce a large number of relations between the scaling dimensions of various composite operators at criticality. These symmetry relations generalize Eq.~(\ref{e1.3}) obtained earlier for the most relevant operators (LDOS moments).
\end{enumerate}

It should be emphasized that we do not attempt to generalize Eq.~(\ref{e1.4}), which is also valid away from criticality, but rather focus on Anderson-transition critical points. The reason is as follows. The derivation of Eq.~(\ref{e1.4}) in Ref.~[\onlinecite{gruzberg11}] was based on a (super-)generalization of a theorem due to Harish-Chandra. We are not able to further generalize this theorem to the non-minimal $\sigma$-models needed for the generalization of Eq.~(\ref{e1.4}) to subleading operators. For this reason, we use a weaker version of the Weyl-invariance argument which is applicable only at criticality. This argument is sufficient to get exact relations between the critical exponents.

In the main part of the paper we focus on the unitary Wigner-Dyson class A. Generalizations to other symmetry classes, as well as some of their peculiarities, are discussed at the end of the paper.

The structure of the paper is as follows. In Sec.~\ref{s2} we briefly review Wegner's classification of composite operators in replica $\sigma$-models by Young diagrams. In Sec.~\ref{s3} we introduce the Iwasawa decomposition for supersymmetric $\sigma$-models of Anderson localization and, on its basis, develop a classification of the composite operators. The correspondence between the replica and SUSY formulations is established in Sec.~\ref{s4} for the case of the minimal SUSY model. Section~\ref{s5} is devoted to the connection between the physical observables (wave-function correlation functions) and the $\sigma$-model composite operators. This subject is further developed in Secs.~\ref{s6a}-\ref{s7}, where we identify observables that correspond to exact scaling operators and thus exhibit pure power scaling (without any admixture of subleading power-law contributions). In Sec.~\ref{s6e} we formulate a complete version (going beyond the minimal-SUSY model considered in Sec.~\ref{s4}) of the
correspondence between the full set of operators of our SUSY classification and the physical observables (wave-function and LDOS correlation functions). An alternative and more basic approach to scaling operators via the notion of highest-weight vector is explained in Sec.~\ref{s8}. We also indicate how this approach is related to the one based on the Iwasawa decomposition. In Sec.~\ref{s9} we employ the Weyl-group invariance and deduce symmetry relations among the anomalous dimensions of various composite operators at criticality. The generalization of these results to other symmetry classes is discussed in Sec.~\ref{s10}. In Sec.~\ref{s11} we analyze the implications of our findings for transport observables defined within the Dorokhov-Mello-Pereyra-Kumar (DMPK) formalism. In Sec.~\ref{s12} we discuss peculiarities of symmetry classes whose $\sigma$-models possess additional O(1) [classes D and DIII] or U(1) [classes BDI, CII, DIII] degrees of freedom. Section \ref{s13} contains a summary of our results,
as well as a discussion of open questions and directions for further research.

\section{Replica $\sigma$-models and\\ Wegner's results}\label{s2}

The replica method leads to the reformulation of the localization problem as a theory of fields taking values in a symmetric space $G/K$ -- a \emph{non-linear $\sigma$-model}. \cite{Wegner79,evers08} If one uses fermionic replicas, the resulting $\sigma$-model target spaces are compact, and for the Wigner-Dyson unitary class (a.k.a.\ class A) they are of the type $G/K$ with $G = \text{U}(m_1 + m_2)$ and $K = \text{U}(m_1) \times \text{U}(m_2)$. Bosonic replicas lead to the non-compact counterpart $G^\prime /K$ where $G^\prime = \text{U}(m_1, m_2)$. The total number of replicas $m = m_1 + m_2$ is taken to zero at the end of any calculation, but at intermediate stages it has to be sufficiently large in order for the $\sigma$-model to describe high enough moments of the observables of interest. The $\sigma$-model field $Q$ is a matrix, $Q = g\Lambda g^{-1}$, where $\Lambda = {\rm diag} (\mathbbm{1}_{m_1}, -\mathbbm{1}_{m_2})$ and $g \in G$. Since $Q$ does not change when $g$ is multiplied on the right ($g \to
gk$) by any element $k \in K$, the set of matrices $Q$ realizes the symmetric space $G/K$. Clearly, $Q$  satisfies the constraint $Q^2 = 1$. Roughly speaking, one may think of the symmetric space $G/K$ as a ``generalized sphere''.

The action functional of the $\sigma$-model has the following structure:
\begin{equation}\label{e2.1}
 S[Q] = \frac{1}{16\pi t} \int \! d^d r \, \text{Tr} (\nabla Q)^2 + h \int \! d^d r \, \text{Tr} Q\Lambda .
\end{equation}
Here $Q({\bf r})$ is the $Q$-matrix field depending on the spatial coordinates $\bf r$. The parameter $1/16\pi t$ in front of the first term is 1/8 times the conductivity (in natural units). In the renormalization-group (RG) framework, $t$ serves as a running coupling constant of the theory. While the first term is invariant under conjugation $Q({\bf r}) \to g_0 Q({\bf r}) g_0^{-1}$ of the $Q$-matrix field by any (spatially uniform) element $g_0 \in G$, the second term causes a reduction of the symmetry from $G$ to $K$, i.e.\ only conjugation $Q({\bf  r}) \to k_0 Q({\bf r}) k_0^{-1}$ by $k_0 \in K$ leaves the full action invariant. The second term provides an infrared regularization of the theory in infinite volume; physically, $h$ is proportional to the frequency. When studying scaling properties, it is usually convenient to work at an imaginary frequency, which gives a non-zero width to the energy levels. If the physical system has a spatial boundary with coupling to metallic leads, then boundary terms arise which are $K$-invariant like the second term in (\ref{e2.1}).

Quite generally, physical observables are represented by composite operators of the corresponding field theory. For the case of the compact $\sigma$-model resulting from fermionic replicas, a classification of composite operators without spatial derivatives was developed by H\"of and Wegner \cite{Hoef86} and Wegner. \cite{Wegner1987a,Wegner1987b} It goes roughly as follows. The composite operators were constructed as polynomials in the matrix elements of $Q$,
\begin{align}\label{e2.2}
  P = \sum_{i_1, \ldots, i_{2k}} T_{i_1,\ldots,i_{2k}} Q_{i_1, i_2} \cdots Q_{i_{2k-1},i_{2k}} .
\end{align}
Such polynomials transform as tensors under the action ($Q \to g Q g^{-1}$) of the group $G = \mathrm{U}(m_1 + m_2)$. They decompose into polynomials that transform irreducibly under $G$, and composite operators (or polynomials in $Q$) belonging to different irreducible representations of the symmetry group $G$ do not mix under the RG flow. The renormalization within each irreducible representation is characterized by a single renormalization constant. Therefore, fixing any irreducible representation it is sufficient to focus on operators in a suitable one-dimensional subspace. In view of the $K$-symmetry of the action, a natural choice of subspace is given by $K$-invariant operators, i.e. those polynomials that satisfy $P(Q) = P(k_0 Q k_0^{-1})$ for all $k_0 \in K$. It can be shown \cite{Hoef86} that each irreducible representation occurring in (\ref{e2.2}) contains exactly one such operator. We may therefore restrict our attention to $K$-invariant operators. By their $K$-invariance, such operators can be represented as linear combinations of operators of the form
\begin{equation}\label{e2.3}
 P_\lambda = \text{Tr} (\Lambda Q)^{k_1} \cdots \text{Tr} (\Lambda Q)^{k_\ell}
\end{equation}
where $\ell = \mathrm{min} \{ m_1 , m_2 \}$, and $\lambda \equiv \{ k_1, \ldots, k_\ell \}$ is a partition $k = k_1 + \ldots + k_\ell$ such that $k_1\ge \ldots \ge k_\ell \geq 0$. In particular, for $k = 1$ we have one such operator, $\{1\}$, for $k = 2 \leq \ell$ two operators, $\{2\}$ and $\{1, 1\}$, for $k = 3 \leq \ell$ three operators $\{3\}$, $\{2, 1\}$, and $\{1, 1, 1\}$, for $k = 4 \leq \ell$ five operators $\{4\}$, $\{3, 1\}$, $\{2, 2\}$, $\{2, 1, 1\}$, and $\{1, 1, 1, 1\}$, and so on. \cite{burmistrov11} As described below, operators of order $k$ correspond to observables of order $k$ in the LDOS or, equivalently, of order $2k$ in the wave-function amplitudes.

It turns out that the counting of partitions yields the number of different irreducible representations that occur for each order $k$ of the operator. More precisely, there is a one-to-one correspondence between the irreducible representations of $G = \mathrm{U}(m_1 + m_2)$ which occur in (\ref{e2.2}) and the set of irreducible representations of $\mathrm{U} (\ell)$, $\ell = \mathrm{min}(m_1,m_2)$, as given by partitions $\lambda = (k_1, \ldots, k_\ell)$. We may also think of the partition $\lambda$ as a Young diagram for $\mathrm{U}(\ell)$. (Please note that Young diagrams and the corresponding partitions are commonly denoted by using parentheses as opposed to the curly braces of the above discussion. An introduction to Young diagrams and their use in our context is given in Appendix \ref{app:notation}.)

The claimed relation with the representation theory of $\mathrm{U}(\ell)$ becomes plausible if one uses the Cartan decomposition $G = KAK$, by which each element of $G$ is represented as $g = k a k'$, where $k,k'\in K$, $a\in A$, and $A \simeq \mathrm{U}(1)^\ell$ is a maximal abelian subgroup of $G$ with Lie algebra contained in the tangent space of $G/K$ at the origin. In this decomposition one has $Q = k a \Lambda a^{-1} k^{-1}$. A $K$-invariant operator $P$ satisfies $P(Q) = P(k a \Lambda a^{-1} k^{-1}) = P(a \Lambda a^{-1})$. In other words, $P$ depends only on a set of $\ell$ ``$K$-radial'' coordinates for $a \in A \simeq \mathrm{U}(1)^\ell$ -- this is ultimately responsible for the one-to-one correspondence with the irreducible representations of $\mathrm{U}(\ell)$.

The $K$-invariant operators associated with irreducible representations are known as zonal spherical functions. For the case of $G/K = \text{U}(2) / \text{U}(1)\times \text{U}(1) = S^2$, which is the usual two-sphere, they are just the Legendre polynomials, i.e.\ the usual spherical harmonics with magnetic quantum number zero; see also Appendix \ref{appendix-hwv}. Please note that here and throughout the paper we use the convention that the symbol for the direct product takes precedence over the symbol for the quotient operation. Thus
\begin{displaymath}
    G / K_1 \times K_2 \equiv G / (K_1 \times K_2).
\end{displaymath}
{}From the work of Harish-Chandra \cite{helgason84} it is known that the zonal spherical functions have a very simple form when expressed by $N$-radial coordinates that originate from the Iwasawa decomposition $G = NAK$; see Sec.~\ref{s3} below. This will make it possible to connect Wegner's classification of composite operators with our SUSY classification, where we use the Iwasawa decomposition.

H\"of and Wegner \cite{Hoef86} calculated the anomalous dimensions of the polynomial composite operators (\ref{e2.2}) for $\sigma$-models on the target spaces $G(m_1 + m_2)/ G(m_1) \times G(m_2)$ for $G = \mathrm{O}$, $\mathrm{U}$, and $\mathrm{Sp}$ (whose replica limits correspond to the Anderson localization problem in the Wigner-Dyson classes A, AI, and AII, respectively) in $2 + \epsilon$ dimensions up to three-loop order. Wegner\cite{Wegner1987a,Wegner1987b} extended this calculation up to four-loop order. The results of Wegner for the anomalous dimensions are summarized in the $\zeta$-function for each composite operator:
\begin{align}\label{e2.4}
 \zeta_\lambda(t) = a_2(\lambda) \rho(t) + \zeta(3) c_3(\lambda) t^4 + O(t^5),
\end{align}
where $t$, serving as a small parameter of the expansion, is the renormalized coupling constant of the $\sigma$-model. The coefficients $a_2$ and $c_3$ depend on the operator $P_\lambda$ (defined by the Young diagram $\lambda$) and on the type of model (O, U, or Sp, as well as $m_1$ and $m_2$). The function $\rho(t)$ depends on the model only and not on $\lambda$. The coefficient $a_2$ happens to be the quadratic Casimir eigenvalue associated to the representation with Young diagram $\lambda$ (\ref{a2}). For the case of unitary symmetry (class A), on which we focus, the coefficients satisfy the following symmetry relations:
\begin{eqnarray}\label{e2.5}
 a_2(\lambda, m) &=& - a_2(\tilde\lambda, - m), \\ c_3(\lambda, m) &=& c_3(\tilde\lambda, -m), \label{e2.6}
\end{eqnarray}
where $\tilde{\lambda}$ is the Young diagram conjugate to $\lambda$, i.e., $\tilde\lambda$ is obtained by reflection of $\lambda$ with respect to the main diagonal. By using the results of Table 2 from Ref.\ [\onlinecite{Wegner1987b}], complementing them with these symmetry relations, and taking the replica limit $m_1 = m_2 = 0$, we can obtain the values of the coefficients $a_2$ and $c_3$ for all polynomial composite operators up to order $k=5$. These values are presented in Table \ref{unitary-table}. The function $\rho(t)$ is given for this model (unitary case, replica limit) by
\begin{equation}\label{e2.7}
 \rho(t) = t + {\textstyle{\frac{3}{2}}}\, t^3 .
\end{equation}
%%%%%%%%%%%%%%%%%%%%%%%%%%%%%%%%%%%%%%%%%%%%%%%%%%%%%%%%%%%%%%%%%%
\begin{table}[t]
  \begin{tabular}{|c|c|c|c|}
    \hline
    $|\lambda|$ & $\lambda$ & $a_2$ & $c_3$ \\
    \hline
    1 & (1)         &  0    & 0     \\
    \hline
    2 & (2)         &  2   & 6     \\
      & (1,1)       &  -2    & 6     \\
    \hline
    3 & (3)         &  6   & 54    \\
      & (2,1)       &  0    & 0     \\
      & (1,1,1)     &  -6    & 54    \\
    \hline
    4 & (4)         &  12  & 216   \\
      & (3,1)       &  4   & 24    \\
      & (2,2)       &  0    & 0     \\
      & (2,1,1)     &  -4    & 24    \\
      & (1,1,1,1)   &  -12   & 216   \\
    \hline
    5 & (5)         & 20   & 600   \\
      & (4,1)       & 10   & 150   \\
      & (3,2)       & 4    & 24    \\
      & (3,1,1)     & 0     & 0     \\
      & (2,2,1)     & -4     & 24    \\
      & (2,1,1,1)   & -10    & 150   \\
      & (1,1,1,1,1) & -20    & 600   \\
    \hline
  \end{tabular}
\caption[]{Coefficients $a_2$ and $c_3$ of the $\zeta$-function for class A in the replica limit. Results for composite operators characterized by Young diagrams up to size $|\lambda| = k = 5$ are shown.}
\label{unitary-table}
\end{table}
%%%%%%%%%%%%%%%%%%%%%%%%%%%%%%%%%%%%%%%%%%%%%%%%%%%%%%%%%%%%%%%%%%%%%%%

% No Young diagrams are shown in this section.
% Move the following comment to a later place?
%
A note on conventions and nomenclature is in order here. The way we draw Young diagrams (see Appendix \ref{app:notation}) is the standard way. Thus the horizontal direction corresponds to symmetrization and the vertical one to antisymmetrization. In Wegner's approach fermionic replicas are used, hence his natural observables are antisymmetrized products of wave functions, whereas the description of symmetrized products (like LDOS moments) requires the symmetry group to be enlarged. Wegner uses a different convention for labeling the invariant scaling operators, employing the Young diagrams conjugate to the usual ones used here. Thus, for example, the LDOS moment $\langle \nu^q \rangle$ corresponds in our convention to the Young diagram $(q)$, while it is labeled by $(1^q)$ in Wegner's works. This has to be kept in mind when comparing our Table \ref{unitary-table} with Table 2 of Ref.\ [\onlinecite{Wegner1987b}]. Of course, if one uses bosonic replicas, the situation is reversed: the natural objects then are symmetrized products and the roles of the horizontal and vertical directions get switched.

While the works [\onlinecite{Hoef86,Wegner1987a,Wegner1987b}] signified a very important advance in the theory of critical phenomena described by non-linear $\sigma$-models, the classification of gradientless composite operators developed there is complete only for compact models. This can be understood already by inspecting the simple example of $\text{U}(2) / \text{U}(1) \times \text{U}(1) = S^2$ (two-sphere), which is the target space of the conventional $\mathrm{O}(3)$ non-linear $\sigma$-model. As was mentioned above, the corresponding $K$-invariant composite operators are the usual spherical harmonics $Y_{l0}$ with $l = 0, 1, \ldots$, which are Legendre polynomials in $\cos\theta$. (The polar angle $\theta$ parametrizes the abelian group $A$, which is one-dimensional in this case.) It is well known that the spherical harmonics indeed form a complete system on the sphere. The angular momentum $l$ plays the role of the size $k = |\lambda|$ of the Young diagram. The situation changes, however, when we
pass to the non-compact counterpart, $\text{U}(1,1)/ \text{U}(1) \times \text{U}(1)$, which is a hyperboloid $H^2$. The difference is that now the polar direction (parametrized by the coordinate $\theta$) becomes non-compact. For this reason, nothing forces the angular momentum $l$ to be quantized. Indeed, the spherical functions on a hyperboloid $H^2$ are characterized by a continuous parameter (determining the order of an associated Legendre function) which takes the role of the discrete angular momentum on the sphere $S^2$. See Appendix \ref{appendix-hwv} for more details.

The above simple example reflects the general situation: for theories defined on non-compact symmetric spaces the polynomial composite operators by no means exhaust the set of all composite operators. In the field theory of Anderson localization, we are thus facing the following conundrum: two theories, a compact and a non-compact one [$\text{U}(m_1 + m_2) / \text{U} (m_1) \times \text{U}(m_2)$ resp.\ $\text{U}(m_1, m_2) / \text{U}(m_1) \times \text{U}(m_2)$ for class A], which should describe in the replica limit $m_1 = m_2 = 0$ the same Anderson localization problem, have essentially different operator content. This is a manifestation of the fact that the replica trick has a very tricky character indeed. In this paper we resolve this ambiguity by using an alternative, well-defined approach based on supersymmetry.
\vfill\eject

\section{SUSY $\sigma$-models: Iwasawa decomposition and classification of composite operators} \label{s3}

In the SUSY formalism the $\sigma$-model target space is the coset space
\begin{align}\label{sigma-model-class-A}
 G/K = \text{U}(n,n|2n)/\text{U}(n|n) \times \text{U}(n|n) .
\end{align}
This manifold combines compact and non-compact features ``dressed'' by anticommuting (Grassmann) variables. Its base manifold $M_0 \times M_1$ is a product of non-compact and compact symmetric spaces: $M_0 = \text{U}(n, n) / \text{U}(n) \times \text{U}(n)$ and $M_1 = \text{U}(2n) / \text{U}(n) \times \text{U}(n)$.

The action functional of the SUSY theory still has the same form (\ref{e2.1}), except that the trace Tr is now replaced by the supertrace STr. It is often useful to consider a lattice version of the model (i.e.\ with discrete rather than continuous spatial coordinates); our analysis based solely on symmetry considerations remains valid in this case as well. Furthermore, it also applies to models with a topological term (e.g., for quantum Hall systems in 2D).

The size parameter $n$ of the supergroups involved needs to be sufficiently large in order for the model to contain the observables of interest; this will be discussed in detail in Sec.~\ref{s5}. The minimal variant of the model with $n = 1$ can accommodate arbitrary moments $\langle \nu^q \rangle$ of the local density of states (LDOS) $\nu$, \cite{gruzberg11} but is in general insufficient to give more complex observables, e.g.\ moments of the Hartree-Fock matrix element (\ref{e1.5}). We will first describe the construction of operators for the $n = 1$ model \cite{gruzberg11} and then the generalization for arbitrary $n$.

Our approach is based on the Iwasawa decomposition for symmetric superspaces, \cite{mmz94,alldridge10} generalizing the corresponding construction for non-compact classical symmetric spaces. \cite{helgason78} The Iwasawa decomposition factorizes $G$ as $G = NAK$, where $A$ is (as above) a maximal abelian subgroup for $G/K$, and $N$ is a nilpotent group defined as follows. One considers the adjoint action (i.e.\ the action by the commutator) of elements of the Lie algebra $\mathfrak{a}$ of $A$ on the Lie algebra $\mathfrak{g}$ of $G$. Since $\mathfrak{a}$ is abelian, all its elements can be diagonalized simultaneously. The corresponding eigenvectors in the adjoint representation are called root vectors, and the eigenvalues are called roots. Viewed as linear functions on $\mathfrak{a}$, roots lie in the space $\mathfrak{a}^*$ dual to $\mathfrak{a}$. A system of positive roots is defined by choosing some hyperplane through the origin of $\mathfrak{a}^*$ which divides $\mathfrak{a}^*$ in two halves, and then defining one of these halves as positive. All roots that lie on the positive side of the hyperplane are considered as positive. The nilpotent Lie algebra $\mathfrak{n}$ is generated by the set of root vectors associated with positive roots; its exponentiation yields the group $N$. The Iwasawa decomposition $G = NAK$ represents any element $g\in G$ in the form $g = nak$, with $n\in N$, $a\in A$, and $k\in K$. This factorization is unique once the system of positive roots is fixed.

An explanation is in order here. The Iwasawa decomposition $G = N A K$ is defined as such only for the case of a non-compact group $G$ with maximal compact subgroup $K$. Now the latter condition appears to exclude the symmetric spaces $G/K$ that arise in the SUSY context, as their subgroups $K$ fail to be maximal compact in general. This apparent difficulty, however, can be circumvented by a process of analytic continuation. \cite{mmz94} Indeed, the classical Iwasawa decomposition $G = N A K$ determines a triple of functions $n : G \to N$, $a : G \to A$, $k : G \to K$ by the uniqueness of the factorization $g = n(g) a(g) k(g)$. In our SUSY context, where $K$ is not maximal compact and the Iwasawa decomposition does not exist, the functions $n(g)$, $a(g)$, and $k(g)$ still exist, but they do as functions on $G$ with values in the complexified groups $N_\mathbb{C}$, $A_\mathbb{C}$, and $K_\mathbb{C}$, respectively. In particular, the Iwasawa decomposition gives us a multi-valued function $\ln a$ which assigns
to every group element $g \in G$ an element $\ln a(g)$ of the (complexification of the) abelian Lie algebra $\mathfrak{a}$.

Note that for $n_0 \in N$, $k_0 \in K$ one has $a(n_0 g k_0) = a(g)$ by construction. Thus one gets a function $\tilde{a}(Q)$ on $G/K$ by defining $\tilde{a}(g \Lambda g^{-1}) \equiv a(g)$. This function is $N$-radial, i.e., it depends only on the ``radial'' factor $A$ in the parametrization $G/K \simeq NA$ and is constant along the nilpotent group $N$: $\tilde{a} (n_0 Q n_0^{-1}) = \tilde{a}(Q)$. Its multi-valued logarithm $\ln \tilde{a}(Q)$ will play some role in what follows.

In the case $n = 1$, which was considered in Ref.\ [\onlinecite{gruzberg11}], the space $\mathfrak{a}^*$ is two-dimensional, and we denote its basis of linear coordinate functions by $x$ and $y$, with $x$ corresponding to the boson-boson and $y$ to the fermion-fermion sector of the theory. In terms of this basis we can choose the positive roots to be
\begin{align}
 2x \,\, (1), && 2iy \,\, (1), && x+iy \,\, (-2), && x-iy \,\, (-2),
\label{positive-roots-class-A-n-1}
\end{align}
where the multiplicities of the roots are shown in parentheses; note that odd roots are counted with negative multiplicity. (A root is called even or odd depending on whether the corresponding eigenspace is in the even or odd part of the Lie superalgebra. Even root vectors are located within the boson-boson and fermion-fermion supermatrix blocks, whereas odd root vectors belong to the boson-fermion and fermion-boson blocks.) For this choice of positive root system, the Weyl co-vector $\rho$ (or half the sum of the positive roots with multiplicities) is
\begin{align}\label{rho-first-choice}
 \rho = -x + iy.
\end{align}

The crucial advantage of using the symmetric-space parametrization generated by the Iwasawa decomposition is that the $N$-radial spherical functions $\varphi_\mu$ have the very simple form of exponentials (or ``plane waves''),
\begin{align}\label{e3.2}
 \varphi_\mu (Q) &= e^{(\rho + \mu)(\ln \tilde{a}(Q))} \cr &= e^{(-1+\mu_0) x(\ln \tilde{a}(Q)) + (1 + \mu_1)i y(\ln \tilde{a}(Q))},
\end{align}
labeled by a weight vector $\mu = \mu_0 x + \mu_1 i y$ in $\mathfrak{a}^*$. The boson-boson component $\mu_0$ of the weight $\mu$ can be any complex number, while the fermion-fermion component is constrained by
\begin{align}\label{e3.3}
 \mu_1 & \in \{-1, -3, -5,\ldots\}
\end{align}
to ensure that $e^{i (1 + \mu_1)(\ln \tilde{a}(Q))}$ is single-valued in spite of the presence of the logarithm.

{}From here on we adopt a simplified notation where we use the same symbol $x$ also for the composition of $x$ with $\ln \tilde{a}$ (and similar for $y$). Thus $x$ may now have two different meanings: either its old meaning as a linear function on $\mathfrak{a}$, or the new one as an $N$-radial function $x \circ \ln \tilde{a}$ on $G/K$. It should always be clear from the context which of the two functions $x$ we mean.

With this convention, Eq.\ (\ref{e3.2}) reads $\varphi_\mu = e^{\rho + \mu} = e^{(-1+\mu_0) x + (1 + \mu_1)i y}$. We will also use the notation
\begin{align}\label{e3.4}
 q &= \frac{1 - \mu_0}{2}, & p = -\frac{1 + \mu_1}{2} \in {\mathbb Z}_+,
\end{align}
where ${\mathbb Z}_+$ means the set of non-negative integers. In this notation the exponential functions (\ref{e3.2}) take the form
\begin{align}\label{plane-wave}
 \varphi_\mu \equiv \varphi_{q,p} = e^{-2qx - 2ipy} .
\end{align}

We mention in passing that the quantization of $p$ is nothing but the familiar quantization of the angular momentum $l$ for the well-known spherical functions on $S^2$. Indeed, the ``momentum'' $p$ is conjugate to the ``radial variable'' $y$ corresponding to the compact (fermion-fermion) sector. The absence of any quantization for $q$ should also be clear from the discussion at the end of Sec.~\ref{s2}. In fact, $q$ is conjugate to the radial variable $x$ of the non-compact (boson-boson) sector, which is a hyperboloid $H^2$.

By simple reasoning based on the observation that $A$ normalizes $N$ (i.e., for any $a_0 \in A$ and $n_0 \in N$ one has $a_0^{-1} n_0 a_0 \in N$), each plane wave $\varphi_\mu$ is an eigenfunction of the Laplace-Beltrami operator and all other invariant differential operators on $G/K$. \cite{eigenfunction} [The same conclusion follows from more general considerations based on highest-weight vectors (see Sec. \ref{s8}).] The eigenvalue of the Laplace-Beltrami operator is
\begin{align}
\label{SUSY-Casimir}
 \mu_0^2 - \mu_1^2 = 4 q(q-1) - 4 p(p+1) ,
\end{align}
up to a constant factor.

It should be stressed that the $N$-radial spherical functions, which depend only on $a$ in the Iwasawa decomposition $g = nak$, differ from the $K$-radial spherical functions (depending only on $a'$ in the Cartan decomposition $g = k'a'k''$, see Sec.~\ref{s2}), since for a given element $Q = g \Lambda g^{-1}$ or $gK$ of $G/K$ the radial elements $a$ and $a'$ of these two factorizations are different. However, a link between the two types of radial spherical function can easily be established. Indeed, if $\varphi(Q)$ is a spherical function, then for any element $k \in K$ the transformed function $\varphi(k g k^{-1})$ is still a spherical function from the same representation. Therefore, we can construct a $K$-invariant spherical function $\tilde{\varphi}_\mu$ by simply averaging $\varphi_\mu(k^{-1} Q k)$ over $K$,
\begin{equation}\label{e3.5}
 \tilde{\varphi}_\mu(Q) = \int_K dk \, \varphi_\mu(k^{-1} Q k) ,
\end{equation}
provided, of course, that the integral does not vanish.

For $n \geq 1$ the space $\mathfrak{a}^*$ has dimension $2n$. We label the linear coordinates as $x_j$, $y_j$ with $j = 1, \ldots, n$; following the notation above, the $x_j$ and $y_j$ correspond to the non-compact and compact sectors, respectively. The positive root system can be chosen as follows:
\begin{align}
 &   x_j - x_k\,\, (2), && x_j + x_k    \,\, (2), && 2x_j \,\, (1), \nonumber \\
 & i(y_l - y_m) \,\, (2), && i(y_l + y_m)\,\, (2), && 2iy_l \,\, (1), \nonumber \\
 & x_j + iy_l \,\, (-2), && x_j - iy_l \,\, (-2), \label{positive-roots-class-A-n}
\end{align}
where $1 \leq j < k \leq n$ and $1 \leq m < l \leq n$. As before, the multiplicities of the roots are given in parentheses, and a negative multiplicity means that the corresponding root is odd, or fermionic. The half-sum of these roots (still weighted by multiplicities) now is
\begin{align}\label{half-sum}
 \rho = \sum_{j=1}^n c_j x_j + i \sum_{l=1}^n b_l y_l
\end{align}
with
\begin{align} \label{e3.6}
 c_j &= 1 - 2j, & b_l &= 2l-1.
\end{align}

The $N$-radial spherical functions are constructed just like for $n = 1$. They are still ``plane waves'' $\varphi_\mu = e^{\rho + \mu}$ but now the weight vector $\mu$ has $2n$ components $\mu^0_j$ and $\mu^1_l$, the latter of which take values
\begin{align}\label{e3.7}
 \mu^1_l & \in \{-b_l, -b_l - 2, -b_l - 4, \ldots\}.
\end{align}
We will also write
\begin{align}\label{e3.8}
 q_j &= -\frac{\mu^0_j + c_j}{2}, & p_l = -\frac{\mu^1_l + b_l}{2} \in {\mathbb Z}_+ .
\end{align}
In this notation our $N$-radial spherical functions are
\begin{align}\label{e3.9}
 \varphi_\mu \equiv \varphi_{q,p} = \exp\Big(-2 \sum_{j=1}^n q_j x_j  - 2i \sum _{l=1}^n p_l y_l \Big).
\end{align}
On general grounds, these are eigenfunctions of the Laplace-Beltrami operator (and all other invariant differential operators) on $G/K$, with the eigenvalue being
\begin{align}\label{e3.10}
 &{\textstyle{\frac{1}{4}}} \sum_{j=1}^n (\mu^0_j)^2 -{\textstyle{\frac{1}{4}}} \sum_{l=1}^n (\mu^1_l)^2 \cr
 &= \sum_{j=1}^n q_j(q_j + c_j) - \sum_{l=1}^n p_l(p_l + b_l) \\
 &= q_1(q_1 - 1) + q_2(q_2 - 3) + \ldots + q_n(q_n - 2n + 1) \cr
 &- p_1(p_1 + 1) - p_2(p_2 + 3) - \ldots - p_n(p_n + 2n - 1) , \nonumber
\end{align}
up to a constant factor.

\section{SUSY--replica correspondence for the $n=1$
  supersymmetric model}\label{s4}

Let us summarize the results of two preceding sections. In Ref.~\ref{s2} we reviewed Wegner's classification of polynomial spherical functions for compact replica models, with irreducible representations labeled by Young diagrams (or sets of non-increasing positive integers giving the length of each row of the diagram). In Sec.~\ref{s3} we presented an alternative classification based on the Iwasawa decomposition of the SUSY $\sigma$-model field. There, the $N$-radial spherical functions are labeled by a set of non-negative integers $p_l$ and a set of parameters $q_j$ that are not restricted to integer or non-negative values. Obviously, the second classification is broader, in view of the continuous nature of the $q_j$. Furthermore, since the SUSY scheme is expected to give, in some sense, a complete set of spherical functions, it should contain Wegner's classification, i.e., each Young diagram of Sec.~\ref{s2} should occur as some $N$-radial plane wave with a certain set of $p_l$ and $q_j$. We are now going
to establish this correspondence explicitly.

\begin{figure}[t]
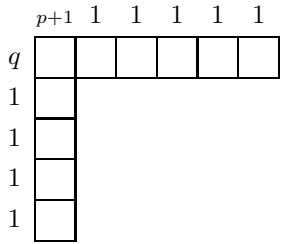

\ytableausetup{mathmode, boxsize=1.5em}
\begin{ytableau}
\none & \none[\scriptstyle p+1] & \none[1] & \none[1] & \none[1] & \none[1] & \none[1] \\
\none[q] & & & & & & \\
\none[1] & \\
\none[1] & \\
\none[1] & \\
\none[1] &
\end{ytableau}
\caption{Hook-shaped Young diagrams $\lambda = (q, 1^p)$ label the scaling operators that can be described within the minimal ($n=1$) SUSY model. The numbers of boxes in each row and in each column are indicated to the left and above the diagram. The example shown in the figure corresponds to $q = 6$, $p = 4$.}
\label{hook}
\end{figure}

We begin with the case of minimal SUSY, $n = 1$.  The starting point is a representation of Green functions as functional integrals over a supervector field containing one bosonic and one fermionic component in both the retarded and advanced sectors. Correlation functions of bosonic fields are symmetric with respect to the spatial coordinates, whereas correlation functions of fermionic fields are antisymmetric. Thus, within the minimal SUSY model one can represent correlation functions involving symmetrization over one set of variables and/or antisymmetrization over another set. On simple representation-theoretic grounds, it follows that the $n=1$ model is sufficient to make for the presence of representations with Young diagrams of the type shown in Fig. \ref{hook}. We refer to such diagrams as {\it hooks} or {\it hook-shaped} for obvious reasons. We introduce two ``dual'' notations (see Appendix \ref{app:notation} for detailed definitions) for Young diagrams by counting the number of boxes either in rows
or in columns; in the first case we put the numbers in round brackets and in the second case in square brackets. In particular, the hook diagram of Fig.\ \ref{hook} is denoted either as $(q,1^p)$ or as $[p+1, 1^{q-1}]$. Below, we point out an explicit correspondence between the spherical functions of the $n=1$ SUSY model and these hook diagrams, by computing the values of the quadratic Casimir operators and identifying the relevant physical observables.

%%%%%%%%%%%%%%%%%%%%%%%%%%%%%%%%%%%%%%%%%%%%%%%%%%%%%%%%%%%%%%%%
%%%%%%%%%%%%%%%%%%%%%%%%%%%%%%%%%%%%%%%%%%%%%%%%%%%%%%%%%%%%%%%%%%%%%

Evaluating the quadratic Casimir (\ref{a2}) for the hook diagrams $(q , 1^p) = [p+1,1^{q-1}]$, and taking the replica limit $m = 0$, we get
\begin{align}\label{e4.1}
 a_2\big((q, 1^p); 0\big) &= q(q-1) - 2 - 4 - \ldots - 2p  \nonumber \\ &= q(q-1) - p(p+1) ,
\end{align}
which is the same (up to a constant factor) as the eigenvalue of the Laplace-Beltrami operator (\ref{SUSY-Casimir}) associated with the plane wave $\varphi_{p,q}$ (\ref{plane-wave}) in the SUSY formalism. This fully agrees with our expectations and indicates the required correspondence: the Young diagram of the type $(q, 1^p)$ of the replica formalism corresponds to the plane wave $\varphi_{q,p}$ (or, more precisely, to the corresponding representation) of the SUSY formalism. While a full proof of the correspondence follows from our arguments in Sec. \ref{s6} (see especially the Sec. \ref{s7}), we feel that the agreement between the quadratic Casimir eigenvalues is already convincing enough for our immediate purposes.

At this point, it is worth commenting on an apparent ``asymmetry'' between $p$ and $q$ in the above correspondence:  the spherical function $\varphi_{q,p}$ corresponds to the Young diagram that has $q$ boxes in its first row but $p+1$ boxes in the first column. The reason for this asymmetry is the specific choice of positive roots (\ref{positive-roots-class-A-n-1}). If instead of $x - iy$ we chose $-x +iy$ to be a positive root (keeping the other three roots), the half-sum $\rho$ would change to
\begin{align}\label{e4.2}
 \tilde \rho = x-iy.
\end{align}
This corresponds to a different choice of nilpotent subgroup $\tilde N$ (generated by the root vectors corresponding to the positive roots) in the Iwasawa decomposition, and thus, to another choice of ($\tilde{N}$-)radial coordinates $\tilde x$ and $\tilde y$ on the superspace $G/K$. As a result, the plane wave
\begin{align}\label{e4.3}
 \varphi_{\tilde q, \tilde p}(\tilde x, \tilde y) = e^{-2{\tilde q}{\tilde x} - 2i{\tilde p}{\tilde y}}
\end{align}
characterized by quantum numbers $\tilde p, \tilde q$ in these new coordinates, is an eigenfunction of the Laplace-Beltrami operator with eigenvalue
\begin{align}\label{e4.4}
 (2{\tilde q} + 1)^2 - (2{\tilde p}-1)^2 &= 4 {\tilde q}({\tilde q} + 1) - 4 {\tilde p}({\tilde p} - 1) .
\end{align}
This is the same eigenvalue as (\ref{SUSY-Casimir}) if one makes the identifications $\tilde{q} = q - 1$ and $\tilde{p} = p + 1$. We thus see that in the new coordinates the asymmetry between $\tilde p$ and $\tilde q$ is reversed: the function $e^{-2{\tilde q}{\tilde x} - 2i{\tilde p}{\tilde y}}$ corresponds to a hook Young diagram with $\tilde{q}+1$ boxes in the first row and $\tilde{p}$ boxes in the first column. Of course the functions $e^{-2qx - 2ipy}$ and $e^{-2{\tilde q}{\tilde x} - 2i{\tilde p}{\tilde y}}$ with $\tilde q = q-1$ and $\tilde p = p+1$ are not identical (since an $N$-radial function is not $\tilde N$-radial in general), but they belong to the same representation.

Choosing a system of positive roots for the Iwasawa decomposition is just a matter of convenience; it is simply a choice of coordinate frame. The positive root system (\ref{positive-roots-class-A-n-1}) is particularly convenient, since with this choice the plane waves corresponding to the most relevant operators (the LDOS moments) depend on $x$ only (and not on $y$). Of course, our final results do not depend on this choice.

We are now going to identify the physical observables that correspond to the operators of the $n=1$ SUSY model. For $p=0$ the Young diagrams of the type $(q, 1^p)$ reduce to a single row with $q$ boxes, i.e.\ $(q) = [1^q]$, which in our SUSY approach represents the spherical function $e^{-2qx}$. As we have already mentioned, this function corresponds to the moment $\langle \nu^q \rangle$ of the LDOS. Note that for symmetry class A, where the global density of states is non-critical, the moment $\langle \nu^q \rangle$ has the same scaling as the expectation value of the $q$-th power of a critical wave-function intensity,
\begin{align}\label{e4.5}
 A_1({\bf r}) = |\psi({\bf r})|^2 .
\end{align}
For the unconventional symmetry classes there is a similarly simple relation; one just has to take care of the exponent $x_\rho$ controlling the scaling of the average density of states, see Eqs.~(\ref{e1.1}) and (\ref{e2.2}). The meaning of the subscript in the notation $A_1$ introduced in Eq.~(\ref{e4.5}) will become clear momentarily. We express the equivalence in the scaling behavior by
\begin{align}
 \langle \nu^q \rangle \sim \langle A_1^q({\bf r}) \rangle.
\end{align}
We now sketch the derivation \cite{gruzberg11} that links $\langle \nu^q \rangle$ with the spherical function $e^{-2qx}$ of the SUSY $\sigma$-model.

The calculation of an observable (i.e.\ some correlation function of the LDOS or of wave functions) in the SUSY approach begins with the relevant combination of Green functions being expressed as an integral over a supervector field. \cite{mirlin94, efetov-book, mirlin-physrep, zirnbauer04} In particular, retarded and advanced Green functions
\begin{equation}\label{e4.5a}
 G_{R,A} ({\bf r}, {\bf r}')  =  (E\pm i\eta - \hat{H})^{-1} ({\bf r}, {\bf r}')
\end{equation}
(where $\eta$ is the level broadening, which for our purposes can be chosen to be of the order of several mean level spacings) are represented as
\begin{eqnarray}\label{e4.6}
 G_R ({\bf r}, {\bf r}') &=& - i \langle S_R({\bf r}) S_R^\ast ({\bf r}')\rangle, \cr G_A ({\bf r}, {\bf r}') &=&  i \langle S_A({\bf r}) S_A^\ast ({\bf r}')\rangle .
\end{eqnarray}
Here $S_{R,A}$ are the bosonic components of the supervector field $\Phi = (S_R,\xi_R,S_A,\xi_A)$ (with subscripts $R,A$ referring to the retarded and advanced subspaces, respectively), and $\langle \ldots \rangle$ on the r.h.s.\ of Eq.~(\ref{e4.6}) denotes the integration over $\Phi$ with the corresponding Gaussian action of $\Phi$. Alternatively, the Green functions can be represented by using the fermionic (anticommuting) components $\xi_{R,A}$; we will return to this possibility below. In order to obtain the $q$-th power $\nu^q$ of the density of states
\begin{equation}\label{e4.7}
 \nu({\bf r}_0) = \frac{1}{2\pi i} \left( G_A({\bf r}_0, {\bf r}_0) - G_R({\bf r}_0, {\bf r}_0) \right) ,
\end{equation}
one has to take the corresponding combination of the bosonic components $S_i$ as a pre-exponential in the $\Phi$ integral:
\begin{align}\label{e4.8}
 \nu^q({\bf r}_0) = \frac{1}{(2\pi)^q q!} \big\langle &\left( S_R({\bf r}_0) - e^{i\alpha} S_A({\bf r}_0) \right)^q \cr \times &\left( S_R^\ast ({\bf r}_0) - e^{-i\alpha} S_A^\ast ({\bf r}_0)\right)^q \big\rangle ,
\end{align}
where $e^{i\alpha}$ is any unitary number. The next steps are to take the average over the disorder and reduce the theory to the non-linear $\sigma$-model form. The contractions on the r.h.s.\ of Eq.~(\ref{e4.8}) then generate the corresponding pre-exponential expression in the $\sigma$-model integral:
\begin{equation}\label{e4.9}
 \langle \nu^q \rangle = 2^{-q} \left\langle \big( Q_{RR} - Q_{AA} + e^{-i\alpha} Q_{RA} - e^{i\alpha} Q_{AR} \big)_{bb}^q \right\rangle,
\end{equation}
where $Q\equiv Q({\bf r}_0)$. The indices $b, f$ refer to the boson-fermion decomposition.

Although the following goes through for any value of $\alpha$, we now take $e^{i\alpha} = 1$ for brevity. It is then convenient to switch to ${\mathcal Q} = Q\Lambda \equiv Q\sigma_3$; here we introduce Pauli matrices $\sigma_j$ in the $RA$ space, with $\sigma_3 = \Lambda$. It is also convenient to perform a unitary transformation ${\mathcal Q} \to \tilde{\mathcal Q} \equiv U \mathcal{Q} U^{-1}$ in the $RA$ space by the matrix $U = (1 + i\sigma_1 + i\sigma_2 + i \sigma_3 ) / 2$, which cyclically permutes the Pauli matrices: $U \sigma_j U^{-1} = \sigma_{j-1}$. The combination of $Q_{ij}$ entering Eq.~(\ref{e4.9}) then becomes
\begin{equation}\label{e4.10}
 (1/2)(Q_{RR} - Q_{AA} + Q_{RA} - Q_{AR})_{bb} = \tilde{\mathcal Q}_{AA,bb} .
\end{equation}
The Iwasawa decomposition $g = nak$ leads to ${\mathcal Q} = n a^2 \sigma_3 n^{-1} \sigma_3$, where we used $k \sigma_3 k^{-1} = \sigma_3$ and $a \sigma_3 a^{-1} = a^2 \sigma_3$. Upon making the transformation ${\mathcal Q} \to \tilde{\mathcal Q}$, this takes the form $\tilde {\mathcal Q} = \tilde{n}\tilde{a}^2\sigma_2\tilde{n}^{-1}\sigma_2$, or explicitly
\begin{equation}\label{e4.11}
 \tilde{\mathcal Q} =
 \begin{pmatrix}
 1 & * & * & * \\
 0 & 1 & * & * \\
 0 & 0 & 1 & * \\
 0 & 0 & 0 & 1
 \end{pmatrix}
 \!\!
 \begin{pmatrix}
 e^{2x} & 0 & 0 & 0 \\
 0 & e^{2iy} & 0 & 0 \\
 0 & 0 & \!e^{-2iy}\! & 0 \\
 0 & 0 & 0 & e^{-2x}
 \end{pmatrix}
 \!\!
 \begin{pmatrix}
 1 & 0 & 0 & 0 \\
 * & 1 & 0 & 0 \\
 * & * & 1 & 0 \\
 * & * & * & 1
 \end{pmatrix},
\end{equation}
where the symbol $*$ denotes some non-zero matrix elements of nilpotent matrices, and we have reversed the boson-fermion order in the advanced sector in order to reveal the meaning of the Iwasawa decomposition in the best possible way. As explained above, the variables $x$ and $y$ parametrize the abelian group $A$ which is non-compact in the $x$-direction and compact in the $y$-direction. By observing that the 44-element of the product of matrices on the r.h.s.\ of (\ref{e4.11}) is $e^{-2x}$, it follows that the matrix element (\ref{e4.10}) is equal to
\begin{align}\label{e4.12}
 \tilde{\mathcal Q}_{AA,bb} = e^{-2x}.
\end{align}
This completes our review of the correspondence between LDOS or wave-function moments and the spherical functions of the SUSY formalism: \cite{gruzberg11}
\begin{align}\label{e4.13}
 \langle A_1^q \rangle \sim \langle \nu^q\rangle \longleftrightarrow
 \varphi_{q,0} = e^{-2qx} .
\end{align}
Let us emphasize that, although our derivation assumes $q$ to be a non-negative integer, the correspondence (\ref{e4.13}) actually holds for any complex value of $q$. Indeed, both sides of (\ref{e4.13}) are defined for all $q \in \mathbb{C}$, and by Carlson's Theorem the complex-analytic function $q \mapsto \langle \nu^q \rangle$ is uniquely determined by its values for $q \in \mathbb{N}$.

At this point the unknowing reader might worry that the positivity of $\langle \nu^q \rangle > 0$ could be in contradiction with the pure-scaling nature of the operator $\varphi_{q,0} = e^{-2qx}$. Indeed, one might argue that if the symmetry group $G$ is compact, then every observable $A$ that transforms according to a non-trivial irreducible representation of $G$ must have zero expectation value with respect to any $G$-invariant distribution.
%
% [cmrz:] Here is the elementary argument:
%
%\begin{displaymath}
%    \langle A_j(gK) \rangle = \langle A_j(g_0 gK) \rangle = \sum_i %\langle A_i(gK) \rangle \rho_{ij}(g_0) .
%\end{displaymath}
%It follows that if $a_j \equiv \langle A_j(gK) \rangle$ were non-zero, %then the vector with components $a_j$ would span a $G$-invariant proper %subspace of the representation space, in immediate contradiction with the %assumption of irreducibility.
%
This apparent paradox is resolved by observing that our symmetry group $G$ is \emph{not} compact (or, if fermionic replicas are used, that the replica trick is very tricky). In fact, the SUSY $\sigma$-model has a non-compact sector which requires regularization by the second term (or similar) in the action functional (\ref{e2.1}). Removing the $G$-symmetry breaking regularization ($h \to 0$) to evaluate observables such as $\langle \nu^q \rangle$, one is faced with a limit of the type $0 \times \infty$ which does lead to a non-zero expectation value $\langle \varphi_{q,0} \rangle \not= 0$.

We now turn to Young diagrams $(1^{\tilde p}) = [\tilde p]$ (where we use the notation ${\tilde p} = p+1$ as before), which encode total antisymmetrization by the permutation group. These correspond to the maximally antisymmetrized correlation function of wave functions. In fact, such a diagram gives the scaling of the expectation value of the modulus squared of the Slater determinant,
\begin{align}\label{e4.14}
 A_{\tilde p}({\bf r}_1, \ldots, {\bf r}_{\tilde p}) &= |D_{\tilde p}({\bf r}_1, \ldots, {\bf r}_{\tilde p})|^2, \\
 D_{\tilde p}({\bf r}_1, \ldots, {\bf r}_{\tilde p}) &= \text{Det} \begin{pmatrix}
 \psi_1({\bf r}_1) & \cdots & \psi_1({\bf r}_{\tilde p}) \\
 \vdots & \ddots & \vdots \\ \psi_{\tilde p}({\bf r}_1) & \cdots & \psi_{\tilde p}({\bf r}_{\tilde p})
 \end{pmatrix}.
\end{align}
Here all points ${\bf r}_i$ are assumed to be close to each other (on a distance scale given by the mean free path $l$), so that after the mapping to the $\sigma$-model they become a single point. Actually, the scaling of the average $\langle A_{\tilde p} \rangle$ with system size $L$ does not depend on the distances $|{\bf r}_i-{\bf r}_j|$ as long as all of them are kept fixed when the limit $L\to\infty$ is taken. However, we prefer to keep the distances sufficiently small, so that our observables reduce to local operators of the $\sigma$-model. Moreover, all of the wave functions $\psi_i$ are supposed to be close to each other in energy (say, within several level spacings). Again, larger energy differences will not affect the scaling exponent; they only set an infrared cutoff that determines the size of the largest system displaying critical behavior. Clearly, $A_{\tilde p}$ reduces to $A_1 = |\psi({\bf r})|^2$ when $\tilde p = 1$ (which was the reason for introducing the notation $A_1$ above).

In the SUSY formalism  the average $\langle A_{\tilde p}\rangle$ can be represented in the following way. We start with
\begin{align}\label{e4.15}
 A_{\tilde p} \sim \big\langle &[\xi_R^*({\bf r}_1)-\xi_A^*({\bf r}_1)]  [\xi_R({\bf r}_1)-\xi_A({\bf r}_1)] \cr \times &[\xi_R^*({\bf  r_2})-\xi_A^*({\bf r}_2)] [\xi_R({\bf r_2})-\xi_A({\bf r}_2)] \ldots \cr
 \times &[\xi_R^*({\bf r}_{\tilde{p}})-\xi_A^*({\bf r}_{\tilde{p}})]
 [\xi_R({\bf r}_{\tilde{p}})-\xi_A({\bf r}_{\tilde{p}})] \big\rangle,
\end{align}
where $\xi, \xi^*$ are the fermionic components of the supervector $\Phi$ used to represent electron Green functions. This expression can now be disorder averaged and reduced to a $\sigma$-model correlation function.
By a calculation similar to that for the moment $\langle \nu^q \rangle$  we now end up with the average of the $\tilde{p}^\mathrm{th}$ moment of the fermion-fermion matrix element $\tilde{\mathcal Q}_{AA,ff}$ of the $\tilde{Q}$ matrix. Here the alternative choice of positive root system mentioned above (for which the radial coordinates were denoted by $\tilde x$, $\tilde y$) is more convenient since, in a sense, it interchanges the roles of $x$ and $y$ in the process of fixing the system of positive roots. As a result, we get the correspondence
\begin{align}\label{e4.16}
 A_{\tilde p} \longleftrightarrow \varphi_{0,\tilde p} =  e^{-2i{\tilde p} {\tilde y}}.
\end{align}

Combining the two examples above [a single-row Young diagram $(q)$ and a single-column Young diagram $(1^{p+1})$], one might guess that a general hook-shaped diagram $(q, 1^p)$ would correspond to the correlator
\begin{align}\label{e4.17}
 \langle A_1^{q-1} A_{p+1} \rangle .
\end{align}
This turns out to be almost correct: the hook diagram $(q, 1^p)$ indeed gives the leading scaling behavior of (\ref{e4.17}). However, the correlation function (\ref{e4.17}) is in general not a pure scaling operator but contains subleading corrections to the scaling for $(q, 1^p)$.

We note in this connection that in the two examples above each of the wave-function combinations $A_1^q$ and $A_p$ corresponds to a single exponential function on the $\sigma$-model target space, and thus to a single $G$-representation. Therefore, at the level of the $\sigma$-model these combinations do correspond to pure scaling operators. (For the LDOS moments $A_1^q$ this was evident from the results of Ref.\ [\onlinecite{gruzberg11}] but we did not stress it there.) We will show below how to construct more complicated wave-function correlators that correspond to pure scaling operators of the $\sigma$-model.

\section{General wave-function correlators}\label{s5}

Clearly, one can construct a variety of wave-function correlators that are different from the totally symmetric ($A_1^q$) and totally antisymmetric ($A_p$) correlators considered in Sec.~\ref{s4}. One example is provided by correlation functions that arise when one studies  the influence of interactions on Anderson and quantum Hall transitions. \cite{burmistrov11} In that context, one is led to consider moments of the Hartree-Fock matrix element (\ref{e1.6}), which involves the antisymmetrized combination (\ref{e1.5}) of two critical wave functions. In terms of the quantities $A_p$ introduced above, Ref.~[\onlinecite{burmistrov11}] calculated
\begin{equation}\label{e5.a1}
 \big\langle A_2(\psi_1,\psi_2; {\bf r}_1, {\bf r}_2) A_2(\psi_3,\psi_4; {\bf r}_3, {\bf r}_4) \big\rangle,
\end{equation}
where the expanded notation indicates the wave functions and corresponding coordinates on which the $A_2$ are constructed. Thus all four points and all four wave functions were taken to be different (although all points and all energies were still close to each other). To leading order, the correlator (\ref{e5.a1}) scales in the same way as  $\langle A_2^2 \rangle$ (where we take $\psi_1 = \psi_3$, $\psi_2 = \psi_4$, ${\bf r}_1 = {\bf r}_3$, ${\bf r}_2 = {\bf r}_4$). The importance of the phrase ``to leading order'' will become clear in Sec.~\ref{s6}.

As we discussed in Sec.~\ref{s4}, the scaling of the average $\langle A_2 \rangle$ is given by the representation with Young diagram  $(1^2) = [2]$. The analysis \cite{burmistrov11} of the second moment $\langle A_2^2 \rangle$ shows that its leading behavior is given by the diagram $(2^2)= [2^2]$. A natural generalization of this is the following proposition: the Young diagram
\begin{align}\label{eq:general-YD-p}
 \lambda = [p_1, p_2, \ldots, p_m]
\end{align}
relates to the replica $\sigma$-model operator that describes the leading scaling behavior of the correlation function
\begin{align}\label{e5.1}
 \langle A_{p_1} A_{p_2} \cdots A_{p_m} \rangle .
\end{align}
We will argue in Sec.\ \ref{s6} below that this is indeed correct. Here we wish to add a few comments.

In general, all combinations $A_{p_i}$ may contain different points and different wave functions (as long as the points and the energies are close) without changing the leading scaling behavior. Thus, a general correlator corresponding to a Young diagram $\lambda$ will involve $|\lambda|$ points and the same number of wave functions. However, if
\begin{equation}
 \lambda = [p_1, p_2, \ldots, p_m] = [k_1^{a_1}, \dots, k_s^{a_s}]
\end{equation}
we may choose to use the same points and wave functions for all $a_i$ combinations $A_{k_i}$ of a given size $k_i$. This yields a somewhat simpler correlator
\begin{align}\label{eq:general-corr-p}
 K_\lambda = \langle A_{k_1}^{a_1} \cdots A_{k_s}^{a_s} \rangle
\end{align}
with the same leading scaling. If we use the alternative notation
\begin{align}\label{eq:general-YD-q}
 \lambda = (q_1, q_2, \ldots, q_n) = (l_1^{b_1}, \dots, l_s^{b_s})
\end{align}
for the Young diagram (\ref{eq:general-YD-p}), the correlator (\ref{eq:general-corr-p}) can also be written as
\begin{align}\label{eq:general-corr-q}
 K_\lambda = \langle A_{b_1}^{l_1 - l_2} A_{b_1 + b_2}^{l_2 - l_3} \cdots
 A_{b_1 + \ldots + b_{s-1}}^{l_{s-1} - l_s} A_{b_1 + \ldots + b_s}^{l_s} \rangle,
\end{align}
see Eqs. (\ref{a-l}) and (\ref{k-b}) in Appendix \ref{app:notation}. In fact, as is easy to see, this can also be rewritten in a natural way as
\begin{align}\label{e1}
 K_{(q_1,\ldots, q_n)} = \langle A_1^{q_1 - q_2} A_2^{q_2 - q_3} \cdots A_{n-1}^{q_{n-1} - q_n} A_n^{q_n} \rangle.
\end{align}
If we introduce the notation
\begin{align}\label{e5.2}
 \nu_1 &= A_1, & \nu_i &= \frac{A_i}{A_{i-1}}, & 2 \leq i \leq n,
\end{align}
then the correlator $K_\lambda$ can also be cast in the following form:
\begin{align}\label{e5.3}
 K_{(q_1,\ldots, q_n)} = \langle \nu_1^{q_1} \nu_2^{q_2} \cdots \nu_{n-1}^{q_{n-1}} \nu_n^{q_n} \rangle.
\end{align}
Below we will establish the correspondence of the correlation functions (\ref{e5.3}) with Young diagrams that was stated in this section. We will also show how to build pure-scaling correlation functions and establish a connection with the Fourier analysis on the symmetric space of the SUSY $\sigma$-model.

\section{Exact scaling operators}\label{s6}

Let us now come back to the issue of exact scaling operators. In the preceding section we wrote down a large family (\ref{e5.1}) of wave-function correlators. In general, the members of this family do not show pure scaling. We are now going to argue, however, that if we appropriately symmetrize (or appropriately choose) the points or wave functions that enter the correlation function, then pure power-law scaling does hold.

%At this point, please be advised of the style of presentation in this %section: results will be stated in a matter-of-fact manner, with numerous %examples given to motivate and make the results plausible. The proof %(actually, a sketch thereof) will be given at the very end.

\subsection{An example}\label{s6a}

Let us begin with the simplest example illustrating the fact stated above. This example is worked out in detail in Sec.~3.3.3 of Ref.~[\onlinecite{mirlin-physrep}] and is provided by the correlation function
\begin{equation}\label{e3}
 \big\langle |\psi_1({\bf r}_1) \psi_2({\bf r}_2)|^2 \big\rangle .
\end{equation}
When the two points and the two wave functions are different, this yields
\begin{equation}\label{e6.1}
 \big\langle (Q_{RR,bb} - Q_{AA,bb})^2  \big\rangle = 2 - 2 \big\langle Q_{RR,bb} Q_{AA,bb} \big\rangle
\end{equation}
after the transformation to the $\sigma$-model. Now the expression $1 - Q_{RR,bb} Q_{AA,bb}$ is not a pure-scaling $\sigma$-model operator: by decomposing it according to representations, one finds that it contains not only the leading term with Young diagram $(2)$, but also the subleading one, $(1,1)$. To get the exact scaling operator for $(2)$, which is
\begin{equation}\label{e4}
 1 -  \big\langle Q_{RR,bb} Q_{AA,bb} + Q_{RA,bb} Q_{AR,bb} \big\rangle,
\end{equation}
one has to symmetrize the product of wave functions in (\ref{e3}) with respect to points (or wave function indices): the correlator that does exhibit pure scaling is
\begin{align}
 \big\langle |\psi_1({\bf r}_1) \psi_2({\bf r}_2) + \psi_1({\bf r}_2) \psi_2({\bf r}_1)|^2 \big\rangle .
\end{align}
Alternatively, one can take the points to be equal and consider the correlation function
\begin{equation}\label{e5}
 \big\langle |\psi_1({\bf r}_1)\psi_2({\bf r}_1)|^2 \big\rangle.
\end{equation}
Then one gets the exact scaling operator right away, since the correlation function (\ref{e5}) already has the required symmetry. One can also take the same wave function:
\begin{equation}\label{e6}
 \big\langle |\psi_1({\bf r}_1)\psi_1({\bf r}_2)|^2 \big\rangle,
 \qquad \big\langle |\psi_1({\bf r}_1)|^4 \big\rangle.
\end{equation}
All of these reduce to the same exact scaling operator (\ref{e4}) in the $\sigma$-model approximation.

\subsection{Statement of result}\label{s6b}

The example above gives us a good indication of how to get wave-function correlators corresponding to pure-scaling operators: the product of wave functions should be appropriately (anti)symmetrized before the square of the absolute value is taken. To be precise, in order to get a pure-scaling correlation function for the diagram (\ref{eq:general-YD-p}) [giving the leading scaling contribution to Eq.~(\ref{e5.1})], one should proceed in the following way:
\begin{itemize}
\item[(i)] View the points and wave functions as filling the Young diagram (\ref{eq:general-YD-p}) by forming the normal Young tableau $T_0$ (see Appendix \ref{app:notation} for definitions);
\item[(ii)] Consider the product of wave-function amplitudes
    \begin{equation}\label{e6.2}
     \psi_1({\bf r}_1) \psi_2 ({\bf r}_2) \cdots \psi_N({\bf r}_N), \quad  N = p_1 + \ldots + p_m.
    \end{equation}
In the notation of Appendix \ref{app:notation} this is $\Psi_\lambda (T_0, T_0)$.
\item[(iii)] Perform the Young symmetrization $c_\lambda = b_\lambda a_\lambda$ according to the rules described in Appendix~\ref{app:notation} (symmetrization $a_\lambda$ with respect to all points in each row followed by antisymmetrization $b_\lambda$ with respect to all points in each column). In this way we obtain
    \begin{equation}\label{e6.2'}
     \Psi_\lambda(T_0, c_\lambda T_0).
    \end{equation}
\item[(iv)] Take the absolute value squared of the resulting expression:
    \begin{equation}\label{e6.2''}
     \big|\Psi_\lambda(T_0, c_\lambda T_0)\big|^2.
    \end{equation}
\end{itemize}

Several comments are in order here. First, one can define several slightly different procedures of Young symmetrization. Specifically, one can perform it with respect to points (as described above) or, alternatively, with respect to wave functions (obtaining $|\Psi_\lambda (c_\lambda T_0, T_0)|^2$). Also, one can perform the Young symmetrization in the opposite order (first antisymmetrization along the columns, then symmetrization along the rows: $\tilde{c}_\lambda = a_\lambda b_\lambda$). In fact, it is not difficult to see that carrying out $\tilde{c}_\lambda$ with respect to wave functions is the same as performing $c_\lambda$ with respect to points, and vice versa, see Eqs.\ (\ref{YS-action-1}), (\ref{YS-action-2}). While for different schemes one will in general obtain from (\ref{e6.2}) slightly different expressions, they will scale in the same way upon averaging, as they belong to the same irreducible representation. Furthermore, once a Young symmetrization of the pro!
 duct (\ref{e6.2}) has been
performed, one can, instead of taking the absolute value squared, simply multiply it with the product $\psi_1^*({\bf r}_1) \psi_2^* ({\bf r}_2) \ldots \psi_N^*({\bf r}_N)$. Finally, the symmetrization with respect to points is redundant if the corresponding points (or wave functions) are taken to be the same, see Eq.\ (\ref{Psi-minimal}).

To illustrate the procedure, let us return again to the correlation function (\ref{e5.a1})
\begin{equation}\label{e2}
 \big\langle A_2(\psi_1,\psi_2; {\bf r}_1, {\bf r}_2) A_2(\psi_3,\psi_4; {\bf r}_3, {\bf r}_4) \big\rangle,
\end{equation}
considered in Ref.~[\onlinecite{burmistrov11}]. As we have already discussed, its leading scaling is that of the Young diagram $(2^2)$; however,\ Eq. (\ref{e2}) includes also corrections due to subleading operators. In order to get the corresponding pure-scaling correlation function, we should start from the product $\psi_1({\bf r}_1)\psi_2({\bf r}_2)\psi_3({\bf r}_3)\psi_4({\bf r}_4)$ and apply the Young symmetrization rules corresponding to the diagram $(2^2)$. This will lead to the expression
\begin{align}\label{e7}
 & [\psi_1({\bf r}_1)\psi_2({\bf r}_2) + \psi_1({\bf r}_2)\psi_2({\bf r}_1)] \cr
 &  \times [\psi_3({\bf r}_3)\psi_4({\bf r}_4) + \psi_3({\bf r}_4)\psi_4({\bf r}_3)],
\end{align}
further anti-symmetrized with respect to the interchange of ${\bf r}_1$ with ${\bf r}_3$ and  with respect to interchange of ${\bf r}_2$ with ${\bf r}_4$. Finally, one should take the absolute value squared. As an alternative to the symmetrization, one can simply set ${\bf r}_1={\bf r}_2$ and ${\bf r}_3={\bf r}_4$ (which means choosing the minimal Young tableau $T_{\text{min}}$ for $T_r$), in which case there is no need to symmetrize. This results in
\begin{align}
 & \big| [\psi_1({\bf r}_1)\psi_3({\bf r}_3) - \psi_1({\bf r}_3)\psi_3({\bf r}_1)] \cr
 &\times [\psi_2({\bf r}_1)\psi_4({\bf r}_3) - \psi_2({\bf r}_3)\psi_4({\bf r}_1)]\big|^2 .
\end{align}
A similar expression can be gotten by setting $\psi_1 = \psi_2$ and $\psi_3 = \psi_4$. Finally, one can do both, keeping only two points and two wave functions. This results exactly in
\begin{align}
 \big|\Psi_{(2^2)}\big(T_{\text{min}}, b_{(2^2)}T_{\text{min}} \big)\big|^2 = \big| D_2^2 \big|^2 = A_2^2,
\end{align}
which is thus a pure scaling operator.

This has a natural generalization to the higher-order correlation functions (\ref{e1}) as follows. Let ${\bf r}_1, \ldots, {\bf r}_n$ be a set of $n$ distinct points. For each $m \leq n$ evaluate $A_m$ at the point ${\bf r}_m$ on a set of wave functions $\psi^{(m)}_1, \ldots, \psi^{(m)}_m$. The coincidence of evaluation points takes care of the symmetrization along all rows of the Young diagram. Moreover, the antisymmetrization is included in the definition of the $A_i$. Therefore, with such a choice of points the correlation function (\ref{e1}) will show pure scaling. This statement is independent of the choice of wave functions $\psi_j^{(m)}$: all of them can be different, or some of them corresponding to different $m$ can be taken to be equal. The most ``economical'' choice is to take only $n$ different wave functions $\psi_1, \ldots, \psi_n$ and for each $m$ set $\psi_j^{(m)} = \psi_j$ (independent of $m$) for $j = 1, \ldots, m$. This is the choice made by the minimal tableau (see Eq.\ (\ref{Psi-minimal-1})):
\begin{align}\label{e6.2'''}
 \Psi_\lambda(T_{\text{min}}, b_\lambda T_{\text{min}}) = D_1^{q_1 -q _2} D_2^{q_2 - q_3} \cdots D_n^{q_n} .
\end{align}
where the numbers $(q_1, \ldots, q_n)$ specify the representation with Young diagram $\lambda$ as in Eq.\ (\ref{eq:general-YD-q}).

\subsection{Sketch of proof}\label{s6c}

We now sketch the proof of the relation between the wave-function correlation functions with the proper symmetry and the $\sigma$-model operators from the corresponding representation. In accordance with Eq.~(\ref{e4.8}), we begin with an integral over a supervector field $S$, \begin{equation}\label{e6.3}
 \big\langle c_\lambda\{S^-_1({\bf r}_1) \cdots S^-_N({\bf r}_N)\}\, c_\lambda\{S^{*-}_1({\bf r}_1)\cdots S^{*-}_N({\bf r}_N)\}\big\rangle.
\end{equation}
As before, $S$ denotes the bosonic components of the superfield; the superscript in $S^-$ reflects the structure in the advanced-retarded space: $S^- = S_R - S_A$. This structure ensures that, upon performing contractions, we get the required combinations of Green functions, $G_R - G_A$. We emphasize, however, that one could equally well choose $S_R + S_A$ or $S_R - e^{i\alpha} S_A$ for any $\alpha$, as was done in Eq.\ (\ref{e4.8}). Indeed, by Eq.\ (\ref{e4.6}) all that matters is that the coefficients of $S_R$ and $S_A$ have the same absolute value. We also mention that the freedom in choosing $\alpha$ is elucidated in more detail in Sec.\ \ref{s8} and Appendix \ref{subsec:hyperboloid}.

The subscript of the $S$-fields in Eq.~(\ref{e6.3}) is the replica index. (Recall that we consider an enlarged number of field components.) The symbol $c_\lambda\{\ldots\}$ denotes the Young symmetrization of the replica indices according to the chosen Young diagram $\lambda = (q_1,q_2, \ldots) = [p_1,p_2, \ldots]$, and $N = |\lambda| = \sum p_i = \sum q_j$. It is given by the product $c_\lambda = b_\lambda a_\lambda$ of the corresponding symmetrization and antisymmetrization operators. (Although in Eq.~(\ref{e6.3}) we put $c_\lambda$ twice, it would actually be sufficient to Young-symmetrize only $S$ fields, or only $S^*$ fields.) It is possible to express the correlation function (\ref{e6.3}) in a more economical way (i.e., by introducing fewer field components), without changing the scaling operator that results on passing to the $\sigma$-model. This economy of description is achieved by observing that
symmetrization is provided simply by the repeated use of the same replica index:
\begin{eqnarray}\label{e6.4}
 && \big\langle b_\lambda \{S^-_1({\bf r}_1^{(1)})\cdots S^-_1({\bf r}_{q_1}^{(1)}) S^-_2({\bf r}_1^{(2)})\cdots S^-_2({\bf r}_{q_2}^{(2)}) \cdots \cr && \times S^-_n({\bf r}_1^{(n)})\cdots S^-_n({\bf r}_{q_n}^{(n)})\} \cr && \times b_\lambda \{S^{*-}_1({\bf r}_1^{(1)}) \cdots S^{*-}_1({\bf r}_{q_1}^{(1)}) S^{*-}_2({\bf r}_1^{(2)})\cdots S^{*-}_2({\bf r}_{q_2}^{(2)}) \cdots \cr && \times S^{*-}_n({\bf r}_1^{(n)})\cdots S^{*-}_n({\bf r}_{q_n}^{(n)})\} \big\rangle.
\end{eqnarray}
Here we denoted by ${\bf r}_1^{(j)}, \ldots, {\bf r}_{q_j}^{(j)}$ the points filling the $j$-th row of the Young diagram $(q_1,\ldots q_n) = [p_1, \ldots, p_m]$, and $b_\lambda\{\ldots\}$ still denotes the operation of antisymmetrization along the columns of the Young diagram.
%
%with respect to each of the  sets of coordinates $\{{\bf r}_1^{(1)}, {\bf %r}_1^{(2)}, \ldots, {\bf r}_1^{(p_1)}\}$ (note that $p_1 = n$), $\{{\bf %r}_2^{(1)}, {\bf r}_2^{(2)}, \ldots\, {\bf r}_2^{(p_2)}\}$, \ldots, %$\{{\bf r}_m^{(1)}, {\bf r}_m^{(2)}, \ldots\, {\bf r}_m^{(p_m)}\}$.
%
Performing all Wick contractions and writing Green functions as sums over wave functions, one sees that Eqs.~(\ref{e6.3}) and (\ref{e6.4}) give (up to an irrelevant overall factor) exactly the Young-symmetrized correlation function of wave functions that was described in Sec.~\ref{s6}. Specifically, the obtained correlation function yields the average of Eq.\ (\ref{e6.2''}).

By the process of transforming to the $\sigma$-model, the $2N$ field values of $S$ and $S^\ast$ in (\ref{e6.3}), (\ref{e6.4}) get paired up in all possible ways to form a polynomial of $N$-th order in the matrix elements of $Q$. The general rule for this is \cite{mirlin-physrep}
\begin{equation}
 S_{p_1}^{-}({\bf r}_1) S^{*-}_{p_2} ({\bf r}_2) \to f(|{\bf r}_1 - {\bf r}_2|) \widehat{\mathcal Q}_{p_1 p_2} \left( {\textstyle{\frac{1}{2}}} ({\bf r}_1 + {\bf r}_2) \right) ,
\end{equation}
where the prefactor $f(|{\bf r}_1 - {\bf r}_2|) = (\pi\nu)^{-1} {\rm Im} \langle G_A({\bf r}_1, {\bf r}_2)\rangle$ depends on the distance between the two points, and $\widehat{\mathcal{Q}} \equiv \tilde{\mathcal{Q}}_{ AA,bb} =
\frac{1}{2} (Q_{RR} - Q_{AA} + Q_{RA} - Q_{AR})_{bb}$ was introduced in Eq.~(\ref{e4.10}). In a 2D system for example, $f(r) = e^{-r/2l} J_0 (k_F r)$. The key properties of the function $f(r)$ are $f(0) = 1$ (more generally, $f(r) \simeq 1$ as long as the distance is much smaller than the Fermi wave length, $r\ll \lambda_F$) and $f(r) \ll 1$ for $r\gg \lambda_F$. In the latter case the corresponding pairing between the fields $S$ and $S^\ast$ can be neglected. Assuming that all points in the correlation function (\ref{e6.3}) are separated by distances $r \gg \lambda_F$, we get an expression of the diagonal structure
\begin{equation}\label{e6.5}
 \big\langle (c_\lambda^{(L)} \otimes c_\lambda^{(R)}) (\widehat{\mathcal Q}_{11} \widehat{\mathcal Q}_{22} \cdots \widehat{\mathcal Q}_{NN}) \big\rangle,
\end{equation}
where $c_\lambda^{(L)} \otimes c_\lambda^{(R)}$ means that we Young-symmetrize separately with respect to \emph{both} sets of indices (left and right). (If in Eq.~(\ref{e6.3}) only one Young symmetrizer is included, then only the corresponding set of indices is Young symmetrized here; this does not change the irreducible representation that Eq.~(\ref{e6.5}) belongs to.) Similarly, starting from Eq.~(\ref{e6.4}) and assuming that all points are sufficiently well separated, we obtain
\begin{equation}\label{e6.6}
 \big\langle (c_\lambda^{(L)} \otimes c_\lambda^{(R)}) (\widehat{\mathcal Q}_{j_1 j_1} \widehat{\mathcal Q}_{j_2 j_2} \cdots \widehat{\mathcal Q}_{j_N j_N}) \big\rangle,
\end{equation}
where the first $q_1$ indices $j_i$ are equal to 1, the next $q_2$ are equal to 2, and so on, and the last $q_n$ are equal to $n$. In this case the operator $a_\lambda$ for symmetrization is redundant (since symmetrization of equal indices has a trivial effect) and we may simplify the expression by replacing $c_\lambda$ by the operator $b_\lambda$ for antisymmetrization along the columns of the Young diagram. One can also take some points in the original expressions (\ref{e6.3}), (\ref{e6.4}) to coincide (provided that the result does not vanish upon antisymmetrization); this will not influence the symmetry and scaling nature of the resulting correlation functions.

To complete our (sketch of) proof, we must show that the polynomial
\begin{equation}
 P_\lambda = (c_\lambda^{(L)} \otimes c_\lambda^{(R)}) (\widehat{\mathcal Q}_{j_1 j_1} \widehat{\mathcal Q}_{j_2 j_2} \cdots \widehat{\mathcal Q}_{j_N j_N})
\end{equation}
is a pure-scaling operator of the non-linear $\sigma$-model. This will be achieved by showing that $P_\lambda$ is an eigenfunction of all Laplace-Casimir operators for $G/K$. The latter can be done in two different ways. Firstly, one may argue with the help of the Iwasawa decomposition that $P_\lambda$ is an $N$-radial spherical function and thus has the desired eigenfunction property. In subsection \ref{s7} below, we spell out this argument along with its natural generalization to complex powers $q$. Secondly, it is possible to get the desired result directly (without invoking the Iwasawa decomposition) by showing that the function $P_\lambda$ is a highest-weight vector for the action of $G$ on the matrices $Q$. This is done in subsection \ref{s8}.

\subsection{Argument via Iwasawa decomposition}\label{s7}

We now argue that the polynomial $P_\lambda$ is an eigenfunction of all Laplace-Casimir operators for $G/K$. To this end, our key observation is that $P_\lambda$ can be written as a product of powers of the principal minors (i.e., in our case, the determinants of the right lower square sub-matrices) of the matrix $\widehat{\mathcal{Q}} \equiv \tilde{\mathcal Q}_{AA,bb}$ for the case of $n$ replicas. Indeed, following the derivation of Eqs.\ (\ref{e6.2'''}), (\ref{Psi-minimal-1}), we can associate the left indices of the $n \times n$ matrix $\widehat{\mathcal Q}$ with one minimal Young tableau, and the right indices with another minimal tableau. As a result, if we denote by $d_j$ the principal minor of $\widehat{\mathcal Q}$ of size $j \times j$, we see that
\begin{align}\label{e6.8}
 P_\lambda \propto d_1^{q_1 -q _2} d_2^{q_2 - q_3} \cdots d_n^{q_n} ,
\end{align}
since the Young symmetrizer $c_\lambda$ here acts essentially as the antisymmetrizer $b_\lambda$, producing determinants of the principal submatrices of $\widehat{\mathcal Q}$.

The final step of the argument is to show that $P_\lambda$ agrees (up to a constant) with the $N$-radial spherical function $\varphi_{q,0}$ of Eq.\ (\ref{e3.9}):
\begin{equation}
 P_\lambda \propto \varphi_{q,0} ,
\end{equation}
which is already known to have the desired property. For that, let us write $\widehat{\mathcal Q}$ in Iwasawa decomposition as
\begin{align}
 \begin{pmatrix} 1 &\ldots &* &* \\ \vdots &\ddots &\vdots &\vdots \\ 0 &\ldots &1 &* \\ 0 &\ldots &0 & 1 \end{pmatrix}
 \begin{pmatrix} e^{-2x_n}\!\!\! &\ldots &0 &0 \\ \vdots & \ddots & \vdots &\vdots \\ 0 & \ldots &e^{-2x_2}\!\!\! &0 \\ 0 &\ldots &0 & e^{-2x_1} \end{pmatrix}
 \begin{pmatrix} 1 &\ldots &0 &0 \\ \vdots &\ddots &\vdots &\vdots \\ * &\ldots &1 &0 \\ * &\ldots &* & 1 \end{pmatrix} .
\end{align}
(Precisely speaking, this is the Iwasawa decomposition of the full matrix $\tilde{\mathcal Q} =\tilde{n}\tilde{a}^2\sigma_2\tilde{n}^{-1}\sigma_2$
projected to the boson-boson part of the right lower block; cf.\ Eq.\ (\ref{e4.11}).) Due to the triangular form of the first and last matrices in this decomposition, the principal minors $d_j$ of this matrix are
\begin{align}
 d_j = \prod_{i=1}^j e^{-2x_i} = \exp\Big(-2\sum_{i=1}^j  x_i\Big).
\end{align}
When this expression is substituted into Eq.\ (\ref{e6.8}), we get exactly the function $\varphi_{q,0}$ of Eq.\ (\ref{e3.9}) for the set $q = (q_1,\ldots,q_n)$ of positive integers $q_j$. Since this function is an eigenfunction of all Laplace-Casimir operators for $G/K$, it follows that $P_\lambda$ has the same property. This completes our proof.

To summarize, recall that in Sec.\ \ref{s6b} we specified a certain set of wave-function correlators. Our achievement here is that we have related these correlators to pure-scaling operators of the non-linear $\sigma$-model. By doing so, we have arrived at the prediction that our wave-function correlators exhibit the same pure-power scaling.

Finally, let us remark that, although the analysis above was formulated in the language of the SUSY $\sigma$-model, it could have been done equally well for the replica $\sigma$-models. (In the presence of a compact sector, where the Iwasawa decomposition is not available without complexification, it would actually be more appropriate to carry out the final step of the argument by the theory of highest-weight vector as outlined in Sec.\ \ref{s8} and Appendix \ref{appendix-hwv}.)

\subsection{Generalization to arbitrary $q_j$}\label{s6e}

We now come to a generalization of our correspondence. The correlators considered in Sec.\ \ref{s6} up to now were polynomials (of even order) in wave-function amplitudes $\psi$ and $\psi^*$, and the resulting $\sigma$-model operators were polynomials in $Q$. The important point to emphasize here is that the wave-function correlation functions (\ref{e1}) are perfectly well-defined for all complex values of the exponents $q_j$ $(j = 1, \ldots, n$). At the same time, while the polynomial $\sigma$-model operators of Wegner's classification clearly require the numbers $q_j$ to be non-negative integers, the $N$-radial spherical functions (\ref{e3.9}) given by the SUSY formalism,
\begin{equation}\label{e6.25}
 \varphi_{q,0} = \exp\Big(-2 \sum_{j=1}^n q_j x_j \Big),
\end{equation}
do exist for arbitrary quantum numbers $q = (q_1, \ldots, q_n)$. Thus one may suspect that our correspondence extends beyond the integers to all values of $q$. This turns out to be true by uniqueness of analytic continuation, as follows.

We gave an indication of the argument in Sec.\ \ref{s4} and will now provide more detail. Let $n = 1$ for simplicity (the reasoning for higher $n$ is no different), and consider
\begin{equation}
    f(q) \equiv \langle \nu^q \rangle / \langle \nu \rangle^q \qquad (q \in \mathbb{C}).
\end{equation}
The triangle inequality gives $| f(q) | \leq f(\mathrm{Re}\, q)$.
By the definition of $\nu$ and the fact that the total density of states is self-averaging, one has an a-priori bound for positive real values of $q :$
\begin{equation}
    0 \leq f(q) \leq (L/l)^{dq} \qquad (q \geq 0),
\end{equation}
where $l$ is the lattice spacing (or UV cutoff) of the $d$-dimensional system. In conjunction with the functional relation \cite{gruzberg11} $f(q) = f(1-q)$, this inequality leads to a bound of the form
\begin{equation}
    | f(q) | \leq e^{C_L (1 + |\mathrm{Re}\,q |)} \qquad (q \in \mathbb{C}),
\end{equation}
where $C_L \propto \ln L$ is a constant. Thus, in finite volume, $f$ is an entire function of exponential type and is also bounded along the imaginary axis. By Carlson's Theorem, this implies that $f$ is uniquely determined by its values on the non-negative integers. It follows that the result of our derivation, taking the pure-scaling correlation functions (\ref{e1}) to $\sigma$-model expectation values of the $N$-radial spherical functions $\varphi_{q,0}$ of (\ref{e6.25}), extends from non-negative integer values of $q$ to all complex values of $q$. This relation is expected to persist in the infinite-volume limit $L \to \infty$.

\subsection{Alternative construction of scaling operators: highest-weight vectors}
\label{s8}

In previous sections we constructed scaling operators in the $\sigma$-model from the Iwasawa decomposition. Here we show how to construct the same operators by using a different approach based on the notion of {\it highest-weight vector}. We just outline the basic idea of this approach, relegating details of the construction to Appendix \ref{appendix-hwv}.

The $\sigma$-model field $Q$ takes values in a symmetric space $G/K$. Our goal is to identify gradientless scaling operators of the $\sigma$-model, i.e., operators that reproduce (up to multiplication by a constant) under transformations of the renormalization group. We know that the change of a local $\sigma$-model operator, say $A$, under an infinitesimal RG transformation can be expressed by differential operators acting on $A$ considered as a function on $G/K$. Assuming that the $\sigma$-model Lagrangian is $G$-invariant, the infinitesimal RG action is by differential operators which are $G$-invariant (also known as Laplace-Casimir operators). Thus a gradientless operator of the $\sigma$-model is a pure scaling operator if it is an eigenfunction of the full set of Laplace-Casimir operators on $G/K$.

Such eigenfunctions can be constructed by exploiting the notion of highest-weight vector, as follows. Let $\mfg \equiv \mathfrak{g}_\mathbb{C}$ denote the {\it complexified} Lie algebra of the Lie group $G$. The elements $X \in \mfg$ act on functions $f(Q)$ on $G/K$ as first-order differential operators $\widehat{X}$:
\begin{align}\label{e8.1}
 (\widehat{X} f)(Q) = \frac{d}{dt}\Big|_{t=0} f \big(e^{-tX} Q \, e^{tX}).
\end{align}
By definition, this action preserves the commutation relations: $[\widehat{X}, \widehat{Y}] = \widehat{[X,Y]}$.

Fixing a Cartan subalgebra $\mfh \subset \mfg$ we get a root-space decomposition
\begin{align}\label{e8.2}
 \mfg = \mfn_+ \oplus \mfh \oplus \mfn_-,
\end{align}
where the nilpotent Lie algebras $\mfn_\pm$ are generated by positive and negative root vectors. We refer to elements of $\mfn_+$ ($\mfn_-$) as raising (resp.\ lowering) operators. (Comparing with the Iwasawa decomposition of Sec.\ \ref{s3}, we observe that $\mathfrak{n}_+$ is the same as the complexification of $\mathfrak{n}$, and $\mathfrak{a}$ is a subalgebra of $\mathfrak{h}$, with the additional generators of $\mathfrak{h}$ lying in the complexified Lie algebra of $K$.)

Now suppose that $\varphi_\lambda$ is a function on $G/K$ with the properties
\begin{align}\label{e8.3}
 &1. \quad \widehat{X} \varphi_\lambda = 0 && \text{for all } X \in \mfn_+, \cr &2. \quad \widehat{H} \varphi_\lambda = \lambda(H) \varphi_\lambda && \text{for all } H \in \mfh.
\end{align}
Thus $\varphi_\lambda$ is annihilated by the raising operators from $\mfn_+$ and is an eigenfunction of the Cartan generators from $\mfh$. Such an object $\varphi_\lambda$ is called a highest-weight vector, and the eigenvalue $\lambda$ is called a highest weight.

Since the Lie algebra acts on functions on $G/K$ by first-order differential operators, it immediately follows that the product $\varphi_{ \lambda_1 + \lambda_2} = \varphi_{\lambda_1} \varphi_{\lambda_2}$ of two highest-weight vectors, as well as an arbitrary power $\varphi_{q\lambda} = \varphi_\lambda^q$ of a highest-weight vector, are again highest-weight vectors with highest weights $\lambda_1 + \lambda_2$ and $q\lambda$, respectively. In the compact case the power $q$ has to be quantized (a non-negative integer) so that $\varphi_\lambda^q$ is defined globally on the space $G/K$. On the other hand, in the non-compact case, we can find a {\it positive} ($\varphi_\lambda > 0$) highest-weight vector, and then it can be raised to an arbitrary complex power $q$.

Now recall that a Casimir invariant $C$ is a polynomial in the generators of $\mfg$ with the property that $[C,X] = 0$ for all $X \in \mfg$. The Laplace-Casimir operator $\widehat{C}$ is the invariant differential operator which corresponds to the Casimir invariant $C$ by the action (\ref{e8.1}). If a function $\varphi_\lambda$ has the highest-weight properties (\ref{e8.3}), then this function is an eigenfunction of all Laplace-Casimir operators of $G$. To see this, one observes that on general grounds every Casimir invariant $C$ can be expressed as
\begin{align}\label{e8.4}
 C = C_\mfh + \sum_{\alpha > 0} D_\alpha X_\alpha,
\end{align}
where every summand in the second term on the right-hand side contains some $X_\alpha \in \mfn_+$ as a right factor. Thus the second term annihilates the highest-weight vector $\varphi_\lambda$. The first term, $C_\mfh$, is a polynomial in the generators of the commutative algebra $\mfh$ and thus has $\varphi_\lambda$ as an eigenfunction by the second relation in (\ref{e8.3}).

In summary, gradientless scaling operators can be constructed as functions that have the properties of a highest-weight vector. To generate the whole set of such operators, one uses the fact that powers and products of heighest-weight vectors are again heighest-weight vectors.

Let us discuss how this construction is related to the Iwasawa decomposition $G = NAK$. $N$-radial functions $f(Q)$ on $G/K$ by definition have the invariance property
\begin{align}\label{e8.5}
 f(n Q n^{-1}) = f(Q) && \forall n \in N.
\end{align}
Any such function is automatically a highest-weight vector if the nilpotent group $N$ is such that its (complexified) Lie algebra coincides with the algebra $\mfn_+$ of raising operators. Indeed, if $X$ is an element of the Lie algebra of $N$, then
\begin{align}\label{e8.6}
 (\widehat{X} f)(Q) = \frac{d}{dt}\Big|_{t=0} f\big(e^{-tX} Q\, e^{tX}) = 0,
\end{align}
since the expression under the $t$ derivative does not depend on $t$ by the invariance (\ref{e8.5}).

In Appendix \ref{appendix-hwv} we implement this construction explicitly. We consider certain linear functions of the matrix elements of $Q$, which we write as
\begin{align}\label{e8.7}
 \mu_Y(Q) = \text{Tr} (YQ).
\end{align}
{}From the definition (\ref{e8.1}) it is easy to see that
\begin{align}\label{e8.8}
 &(\widehat{X} \mu_Y )(Q) = \frac{d}{dt}\Big|_{t=0} \text{Tr} \big(e^{tX} Y e^{-tX} Q) = \mu_{[X,Y]}(Q).
\end{align}
Then, if $[X,Y] = 0$, the function $\mu_Y(Q)$ is annihilated by $\widehat{X}$. To construct highest-weight vectors, which are annihilated by all $\widehat{X}$ for $X \in \mfn_+$, we then build certain polynomials from these linear functions, and form products of their powers. In this manner we recover exactly the set of scaling operators (\ref{e6.8}), (\ref{e6.25}).

\section{Weyl group and symmetry relations between scaling exponents}
\label{s9}

In the preceding sections we constructed wave-function observables that show pure-power scaling, by establishing their correspondence with scaling operators of the SUSY $\sigma$-model. Now we are ready to explore the impact of Weyl-group invariance on the spectrum of scaling exponents for these operators (and the corresponding observables) at criticality. The Weyl group $W$ is a discrete group acting on the Lie algebra $\mathfrak{a}$ of the group $A$, or equivalently, on its dual $\mathfrak{a}^*$. Acting on $\mathfrak{a}^*$, $W$ is generated by reflections $r_\alpha$ at the hyperplanes orthogonal to the even roots $\alpha$:
\begin{equation}\label{e9.1}
 r_\alpha: \; \mathfrak{a}^* \to \mathfrak{a}^*, \quad \mu \mapsto \mu
 - 2\alpha \frac{\langle \alpha,\mu\rangle}{\langle\alpha,\alpha\rangle},
\end{equation}
where $\langle\cdot,\cdot\rangle$ is the Euclidean scalar product of the Euclidean vector space $\mathfrak{a}^\ast$.

Key to the following is the Harish-Chandra isomorphism, see Refs.~[\onlinecite{helgason78}, \onlinecite{helgason84}] for the classical version and Ref.~[\onlinecite{alldridge10}] for the SUSY generalization (which we need). The statement is that there exists a homomorphism (actually, an isomorphism in classical situations) from the algebra of $G$-invariant differential operators on $G/K$ to the algebra of $W$-invariant differential operators on $A$. This homomorphism (or isomorphism, as the case may be) is easy to describe: given a $G$-invariant differential operator $D$ on $G/K$, one restricts it to its $N$-radial part, which can be viewed as a differential operator on $A$, and then performs a so-called Harish-Chandra shift $(\lambda \to \lambda - \rho)$ by the half-sum of positive roots $\rho$. The shifted operator turns out to be $W$-invariant.

This property of $W$-invariance is what matters to us here, for it has the consequence that if $\chi_\mu(D)$ denotes the eigenvalue of $D$ on the spherical function (or highest-weight vector) $\varphi_\mu$, see Eq.\ (\ref{e3.9}), then
\begin{equation}\label{e9.3}
 \chi_{w\mu} = \chi_\mu
\end{equation}
for all $w \in W$. In words: if two spherical functions $\varphi_\mu$ and $\varphi_\lambda$ have highest weights $\lambda = w \mu$ related by a Weyl-group element $w \in W$, then their eigenvalues are the same, $\chi_\mu(D) = \chi_\lambda(D)$, for any $D$. To the extent that the $\sigma$-model renormalization group transformation is $G$-invariant (and hence is generated by some $G$-invariant differential operator on $G/K$), we have the following important consequence: the scaling dimensions of the scaling operators (which arise as eigenvalues of the $G$-invariant operator associated with the fixed point of the RG flow) are $W$-invariant.

For our purposes it will be sufficient to focus on the subgroup of the Weyl group which is generated by the following transformations on $\mathfrak{a}^\ast$: (i) sign inversion of any one of the $\mu^0$-components: $\mu_i^0 \to - \mu_i^0$ (reflection at the hyperplane $\mu_i^0 = 0$), and (ii) pairwise exchange of $\mu^0$-components: $\mu_i^0 \leftrightarrow \mu_j^0$ (reflection at the hyperplane $\mu_i^0-\mu_j^0 =0 $). In view of Eq.~(\ref{e3.8}) these induce the following transformations of the plane-wave numbers $q_j$:
\begin{enumerate}
 \item[(i)] sign inversion of $q_j + \dfrac{c_j}{2}$ for any $j\in \{1, 2, \ldots, n \}$:
\begin{equation}\label{e9.4}
 q_j \to - c_j - q_j,
\end{equation}
where $c_j$ is the coefficient in front of $x_j$ in the expression for the half-sum $\rho$ of positive roots, see Eqs.~(\ref{half-sum}), (\ref{e3.6});
\item[(ii)] permutation of $q_i + \dfrac{c_i}{2}$ and $q_j + \dfrac{c_j}{2}$ for some pair $i, j \in \{1, 2, \ldots, n \}$:
\begin{equation}\label{e9.5}
 q_i \to q_j + \frac{c_j - c_i}{2}; \quad q_j \to q_i + \frac{c_i - c_j}{2} .
\end{equation}
\end{enumerate}

By combining all such operations, one generates a subgroup $W_0 \subset W$ of the Weyl group. Whenever two scaling operators with quantum numbers $q = (q_1,q_2,\ldots,q_n)$ and $q' = (q_1',q_2',\ldots,q_n')$ are related by a Weyl transformation $w \in W_0$, the scaling dimensions of these scaling operators must be equal. We now present some examples of this general statement. As before, we focus on class A, for which $c_j = 1-2j$, see Eq.~(\ref{e3.6}). Generalizations to the other classes will be discussed below.

Consider first the most symmetric representations $(q)$, which are characterized by a single number $q_1 \equiv q$. (Here, for convenience, we continue to use Young-diagram notation, even though $q_1$ need not be a positive integer and does not correspond to a representation of polynomial  type.) The invariance under the Weyl group then implies that the following two representations
\begin{equation} \label{e9.6}
 (q), \quad (1-q),
\end{equation}
(here we used $c_1 = - 1$) give identical scaling dimensions. This is exactly the symmetry statement (\ref{e1.3}) governing the multifractal scaling of the LDOS moments.

Next, consider representations of the type $(q_1,q_2)$. By applying the Weyl symmetry operations above, we can generate from it a series of 8 representations:
\begin{eqnarray}\label{e9.7}
 && (q_1,q_2),\ \ (1-q_1,q_2),\ \ (q_1,3-q_2),\ \ (1-q_1,3-q_2),
 \nonumber \\ && (2-q_2,2-q_1),\ \ (-1+q_2,2-q_1), \nonumber \\
 && (2-q_2,1+q_1),\ \ (-1+q_2,1+q_1).
\end{eqnarray}
Again, all of them are predicted to give the same scaling dimension. As an important example, starting from the trivial representation $(0) \equiv (0,0)$ (i.e.\ the unit operator) we generate the following set:
\begin{eqnarray}\label{e9.8}
&& (0,0)\,,\ \ (1,0)\,,\ \ (0,3)\,,\ \ (1,3)\,, \nonumber \\
&& (2,2)\,,\ \ (-1,2)\,,\ \ (2,1)\,,\ \ (-1,1)\,.
\end{eqnarray}
Since $(0,0)$ has scaling dimension zero, we expect the same to hold for all other representations of this list -- as long as the Anderson-transition fixed point is in class A. This is a remarkable statement.

In fact, among the set of representations (\ref{e9.8}), four are of polynomial type and ``standard'' in that they are also present in the replica approach of Wegner. Aside from the trivial representation $(0,0)$, these are $(1,0)$, $(2,2)$, and $(2,1)$. The representation $(1,0)$ corresponds to $\langle Q\rangle$ which is well known to be non-critical in the replica limit. However, for the polynomial representations $(2,2)$ and $(2,1)$, our exact result seems to be new. It is worth emphasizing that Wegner's four-loop perturbative $\zeta$-function \cite{Wegner1987b} is fully consistent with our finding: both coefficients $a_2$ and $c_3$ vanish for these operators in the replica limit, see Table~\ref{unitary-table} above. Moreover, a numerical analysis \cite{burmistrov11} of the correlation function (\ref{e5.a1}), whose leading scaling behavior is controlled by $(2,2)$, also yielded a result consistent with $\chi_{(2,2)} = 0$.

For our next example of importance, consider the case of $q_1 = q_2$. Inspecting Eq.~(\ref{e9.7}) we see, in particular, that the $\sigma$-model operators $(q,q)$ and $(2-q,2-q)$ have the same scaling dimensions. Now we know [see Sec.~\ref{s7} and Eq.~(\ref{e1})] that the operator $(q,q)$ corresponds to the moment $\langle A_2^q \rangle$ of the Hartree-Fock type correlation function $A_2$, Eq.\ (\ref{e1.5}). Thus we learn that the
multifractal spectrum of scaling dimensions for the Hartree-Fock moments $A_2^q$ is symmetric under the reflection $q \leftrightarrow 2-q$.

One can continue these considerations and look at equivalences between scaling dimensions for $n = 3$, i.e.\ for operators $(q_1,q_2,q_3)$, and so on. In general, the Weyl orbit of an operator $(q_1,\ldots,q_n)$ with $n$ different components consists of $2^n n!$ operators (due to $2^n$ sign inversions and $n!$ permutations) with equal scaling dimensions. We have checked that all results obtained by Wegner, \cite{Wegner1987b} who analyzed operators described by Young diagrams up to size 5 and up to four loops (see Table~\ref{unitary-table}), are fully consistent with this prediction.

This includes the afore-mentioned representations $(1)$, $(2,1)$, and $(2,2)$, as well as $(3,1,1)$: all of them are related to the trivial representation by Weyl-group operations and, indeed, Wegner obtained zero values of $a_2$ and $c_3$ for all of them (in the replica limit). Furthermore, the operators $(3,2) = [2,2,1]$ and $(3,1) = [2,1,1]$ are clearly related to each other by the Weyl reflection $q_2 \to 3-q_2$. Again, as is shown in Table~\ref{unitary-table}, Wegner's four-loop results, $a_2 = 4$ and $c_3 = 24$, are the same for these.

\section{Other symmetry classes}\label{s10}

In order to apply the Weyl-symmetry argument to the other symmetry classes, we need the expressions for the half-sum of positive roots for them. More specifically, we will now present the ``bosonic'' part $\rho_b$ (which is a linear combination of the basic functions $x_j$) of $\rho$. By transcription of the above analysis, its coefficients $c_j$ determine the Harish-Chandra shift entering the Weyl transformation rules for the operators $(q_1,\ldots,q_n)$, see Eqs.~(\ref{e9.4}) and (\ref{e9.5}).

The root systems for all symmetry classes are listed in Table~\ref{table:root-systems} of Appendix \ref{appendix-B}. The resulting  $\rho_b$ are
\begin{equation}\label{e10}
 \rho_b = \sum c_j x_j ,
\end{equation}
where the coefficients $c_j$ ($j = 1, 2, \dots$) read
\begin{align}
 &c_j = 1-2j, &&  \text{class A}, \label{e9a} \\
 &c_j = -j, &&  \text{class AI}, \label{e9b} \\
 &c_j = 3-4j, &&  \text{class AII}, \label{e9c} \\
 &c_j = 1-4j, &&  \text{class C},  \label{e9d}\\
 &c_j = 1-j, &&  \text{class D},  \label{e9e}\\
 &c_j = -2j, &&  \text{class CI},  \label{e9f}\\
 &c_j = 2-2j, &&  \text{class DIII}, \label{e9g} \\
 &c_j = \frac{1}{2}-j, &&  \text{class BDI}, \label{e9h} \\
 &c_j = 2-4j, &&  \text{class CII}, \label{e9i} \\
 &c_j = 1-2j, &&  \text{class AIII}.  \label{e9j}
\end{align}

The results obtained above for class A generalize in a straightforward manner to four of the other classes, which comprise the two remaining Wigner-Dyson classes, AI and AII, and two of the Bogoliubov-de Gennes classes, C and CI. The Weyl-symmetry operations involve the pertinent values of $c_j$ in each case. For example, for the most symmetric operators $(q)$ (characterizing the LDOS moments) we obtain the correspondence $(q) \leftrightarrow (-c_1-q)$, where $-c_1$ has value 1 for the classes A, AI, and AII, value 2 for class CI, and value 3 for class C. This is exactly the symmetry (\ref{e1.3}) obtained in Ref.~[\onlinecite{gruzberg11}], with $q_* = - c_1$.

Correspondences between the representations with two or more numbers $(q_1, \ldots, q_n)$ are obtained in exactly the same way as described in Sec.~\ref{s9} for class A. Again, we have checked that the four-loop results of Wegner \cite{Wegner1987b} for the orthogonal and symplectic classes (AI and AII) conform with our exact symmetry relations. Specifically, for class AI, our results imply the following Weyl-symmetry relations (and thus equal values of the scaling dimensions): (i) $(2,2) \leftrightarrow (2)$; (ii) $(1,1) \leftrightarrow (2,1,1)$; (iii) $(3,2) \leftrightarrow (3)$; (iv) $(2,2,1) \leftrightarrow (1) \leftrightarrow (0)$ (scaling exponent equal to zero); these are the Young diagrams up to size $5$ studied in Ref.~[\onlinecite{Wegner1987b}]. For class AII the dual correspondences hold: (i) $(2,2) \leftrightarrow (1,1)$; (ii) $(2) \leftrightarrow (3,1)$; (iii) $(2,2,1) \leftrightarrow (1,1,1)$; (iv) $(3,2) \leftrightarrow (1) \leftrightarrow (0)$. Needless to say, the results of Ref.~[\onlinecite{Wegner1987b}] for the coefficients $a_2$, $c_3$ for these operators do conform with the predicted relations.

Generalization to the remaining five classes (D, DIII, BDI, CII, and AIII) is more subtle due to peculiarities of their $\sigma$-model manifolds. We defer this issue to Sec.~\ref{s12}.

\section{Transport observables}\label{s11}

We now address the question whether the classification and symmetry analysis of the present paper are also reflected in transport observables. To begin, we remind the reader that such a correspondence between wave-function and transport observables has previously been found for the case of the $(q)$ operators. Specifically, one can consider the scaling of moments of the two-point conductance at criticality, \cite{janssen99,klesse01}
\begin{equation}\label{e11.1}
 \langle g^q({\bf r},{\bf r}')\rangle \sim |{\bf r}-{\bf r}'|^{ -\Gamma_q}.
\end{equation}
Actually, $g^q({\bf r},{\bf r}')$ is not a pure-scaling operator \cite{janssen99} (unlike the LDOS moments considered above), thus Eq.~(\ref{e11.1}) should be understood as characterizing the leading long-distance behavior of $\langle g^q({\bf r},{\bf r}')\rangle$. Nevertheless, it turned out that the transport exponents $\Gamma_q$ and the LDOS exponents $x_q = \Delta_q + qx_\rho$ are related as \cite{evers08}
\begin{equation}\label{e17}
 \Gamma_q = \left \{ \begin{array}{ll} 2x_q, & \qquad q\le q_*/2, \\
 2x_{q^*}, & \qquad q \ge q_*/2 . \end{array} \right.
\end{equation}
Notice that while the LDOS spectrum $x_q$ is symmetric with respect to the point $q_*/2 = -c_1/2$, the two-point conductance spectrum $\Gamma_q$ ``terminates'' (i.e., has a non-analyticity and becomes constant) at this point. Yet, the spectrum $\Gamma_q$ clearly carries information about the Weyl symmetry: if one performs its analytic continuation (starting from the region below $q_*/2$), one gets the spectrum $2x_q = 2x_{q^\ast - q}$.

%{\color{blue}
A physically intuitive argument explaining Eq.~(\ref{e17}) is as follows. For sufficiently low $q$, the moments of $g({\bf r},{\bf r}')$ are controlled by small values of the conductance. When $g({\bf r},{\bf r}')$ is small, one can think of it as a tunneling conductance that is proportional to the product of the LDOS at the points ${\bf r}$ and ${\bf r'}$. The corresponding correlation function $\langle \nu^q({\bf r}) \nu^q({\bf r'}) \rangle$ scales with $|{\bf r-r'}|$ with the exponent $2x_q$, in agreement with the first line of Eq.~(\ref{e17}). (This argument can be also cast in the RG language, see the end of Sec.~\ref{s11.2}.) On the other hand, the two-point conductance cannot be larger than unity. For this reason the relation $\Gamma_q = 2 x_q$ does not hold beyond the symmetry point $q = q_*/2$. The moments with $q\ge q_*/2$ are controlled by the probability to have  $g({\bf r},{\bf r}')$ of order unity.

In view of the relation  (\ref{e17}), a natural question is whether there are any transport observables corresponding to the composite operators $(q_1,q_2,\ldots)$ beyond the dominant one, $(q)$. We argue below that this is indeed the case, construct explicitly these transport observables, and conjecture a relation between the critical exponents.
%}

In order to get some insight into this problem, it is instructive to look first at quasi-1D metallic systems, whose transport properties can be described within the DMPK formalism. \cite{beenakker97, evers08} The rationale behind this is as follows. First, the classification of transport observables that we are aiming at is based (in analogy with the classification of wave-function observables as developed above) purely on symmetry considerations and, therefore, should be equally applicable to metallic systems. Second, a 2D metallic system is ``weakly critical'' (at
distances shorter than the localization length), and the corresponding anomalous dimensions can be studied within the perturbative RG (which is essentially the same as Wegner's RG analysis in $2 + \epsilon$ dimensions). By a conformal
mapping, a 2D system can be related to the same problem in a quasi-1D geometry (with a power-law behavior translating into an exponential decay). Therefore, if some symmetry properties of spectra of transport observables generically hold at criticality, we may expect to see some manifestations of them already in the solution of the DMPK equation.

\subsection{DMPK, localized regime}\label{s11.1}

In the DMPK approach, the transfer matrix of a quasi-1D system is described by ``radial'' coordinates (w.r.t.\ a Cartan decomposition) $X_j$, $j = 1, 2, \ldots, N$, where $N$ is the number of channels. All transport properties of the wire are expressed in terms of these radial coordinates. In particular, the dimensionless conductance is
\begin{equation}\label{e11.2}
 g = \sum_{j=1}^N T_j = {\rm Tr} \: T = {\rm Tr}\: tt^\dagger\,,
\end{equation}
where
\begin{equation}\label{e11.3}
 T_j = \frac{1}{\cosh^2 X_j}
\end{equation}
are the transmission eigenvalues, i.e.\ the eigenvalues of $T = tt^\dagger$ (and $t^\dagger t$), where $t$ is the transmission matrix.

The DMPK equations describe the evolution with system length (playing the role of a fictitious time) of the joint distribution function for the transmission eigenvalues (or the coordinates $X_j$), and they have the form of diffusion equations on the symmetric space associated with the noncompact group of transfer matrices. In the localized regime, where the wire length $L$ is much larger than the localization length $\xi$, the typical value of each transmission eigenvalue becomes exponentially large relative to the next one: $1 \gg T_1 \gg T_2 \gg \ldots \gg T_N$. As a result, the equations for the random variables $X_j$ decouple, yielding an advection-diffusion equation for each $X_j$. The solution has a Gaussian form, with both the average $\langle X_j \rangle$ and the variance ${\rm var}(X_j)$ proportional to $L/\xi$ and with ${\rm var}(X_j)$ independent of $j$. Each of the symmetry classes therefore gives rise to a set of numbers $\langle X_j \rangle / {\rm var}(X_j)$ (which depend solely on the
corresponding symmetric spaces). Remarkably, comparing the above results (\ref{e9a})--(\ref{e9j}) with the known DMPK results, we observe that for all symmetry classes one has
\begin{equation}\label{e11.4}
 - c_j = \frac{\langle X_j \rangle}{{\rm var}(X_j)} .
\end{equation}
(In the case of the chiral classes, we note that Eq.~(\ref{e11.4}) holds when the $X_j$ evolve according to the DMPK equations with an even number of channels.)

This result allows us to draw a link between the transport quantity $T_j$ and the LDOS observable $\nu_j$ defined in Eq.~(\ref{e5.2}). Indeed, if we use the approximation $T_j \approx 4e^{-2X_j}$, which is valid in the localized regime, we get
\begin{equation}\label{e11.5}
 \langle T_j^q \rangle \sim \exp\{2v q(q+c_j)\}, \qquad v = {\rm var}(X_j).
\end{equation}
This expression for $\langle T_j^q \rangle$ has a point $q = -c_j/2$ of reflection symmetry. [We should add that this requires a continuation of Eq.~(\ref{e11.5}) from its range of actual validity to a region of larger $q$, see the discussion below Eq.~(\ref{e17}).] Now we recall that the scaling of $\langle \nu_j^q \rangle$ is determined by the representation $(0,\ldots,0,q,0,\ldots)$, with $q$ at the $j$-th position, see Eq.~(\ref{e5.3}). Hence $\langle \nu_j^q\rangle \sim \langle \nu_j^{ -c_j-q}\rangle$, i.e.\ the symmetry point of the multifractal spectrum for $\nu_j$ is exactly $-c_j/2$. This links $T_j$ with $\nu_j$, as stated above.

We now write
\begin{align}
\label{e15}
T_m &= \frac{T_1 T_2 \cdots T_m}{T_1 T_2 \cdots T_{m-1}} = \frac{S_m}{S_{m-1}},  & S_m =T_1 \cdots T_m,
\end{align}
and draw an analogy between Eq.~(\ref{e15}) and Eq.~(\ref{e5.2}). Specifically, $T_m$ corresponds to $\nu_m$ (as we have already seen earlier) and $S_m$ to $A_m$. To further strengthen the analogy, we point out that $S_m$ can be presented in the form of the absolute value squared of a determinant. Indeed, consider first $m=2$. Choose two incoming ($p,q$) and two outgoing ($r,s$) channels and consider the $2 \times 2$ matrix $t^{(2)}$ formed
by the elements $t_{ij}$, $i=p,q$, $j=r,s$, of the transmission matrix. Then calculate the absolute value squared of the determinant of this matrix, and sum over the channel indices $p,q,r,s$:
\begin{align}
\label{e16}
&\sum_{p,q,r,s} \big|{\rm det}\:t^{(2)}_{ij} \big|^2 = \sum_{p,q,r,s} |t_{pr}t_{qs} - t_{ps}t_{qr}|^2
\nonumber \\
&= 2 ({\rm Tr} \, tt^\dagger)^2 - 2 {\rm Tr} (tt^\dagger)^2 = 2 ({\rm Tr} \, T)^2 - 2 {\rm Tr} \, T^2
\nonumber \\
&= 2\big[(T_1 + T_2)^2 - (T_1^2 + T_2^2)\big] = 4 T_1 T_2 = 4S_2.
\end{align}

The same applies to higher correlation functions: by considering the determinant of an $m\times m$ matrix $t^{(m)}$ and taking its absolute value squared, we get $S_m$ (up to a factor). If the total number of channels is $m$, this is straightforward (the modulus squared of the determinant then equals $T_1 T_2 \cdots T_m$); if the total number of channels is larger than $m$, then, strictly speaking, averaging over the choice of $m$ channels is required. We expect, however, that the determinant will typically behave in the same way for any choice of the channels.

To summarize, the transmission eigenvalues $T_m$ of the DMPK model characterize the leading contribution to the decay of transport quantities $S_m/S_{m-1}$, where the $S_m$ are given by the absolute values squared of the determinants of $m\times m$ transmission matrices. There is a clear correspondence between the wave-function observables $\nu_i = A_i/A_{i-1}$ and the transport observables $T_m = S_m/S_{m-1}$. In the next subsection we generalize this construction to critical systems.

\subsection{Transport observables at criticality}\label{s11.2}

We are now ready to formulate a conjecture about the scaling of subleading transport quantities at criticality. It generalizes the relation (\ref{e17}) between the scaling exponents of the moments of the conductance ($\Gamma_q$) and of the LDOS ($x_q$).

Consider a system at criticality and take two points ${\bf r}_1$ and ${\bf r}_2$ separated by a (large) distance $R$. Attach $N$ incoming and $N$ outgoing transport channels near each of these two points. This yields an $N\times N$ transmission matrix $t$. Define $B_m$ as the absolute value squared of the determinant of its upper-left $m\times m$ corner (i.e., of the transmission matrix $t^{(m)}$ for the first $m$ incoming and first $m$ outgoing channels). This lets us build a family of transport correlation functions ($n\le N$):
\begin{eqnarray}\label{e18}
 M_{q_1 q_2 \ldots q_n}(R) &=& \big\langle B_1^{q_1-q_2} B_2^{q_2-q_3} \cdots B_{n-1}^{q_{n-1}-q_n}B_n^{q_n} \big\rangle \nonumber \\
 & = & \langle \tau_1^{q_1} \cdots \tau_n^{q_n}\rangle,
\end{eqnarray}
where $\tau_n = B_n / B_{n-1}$. The conjecture is that the critical index $\Gamma_{q_1 q_2 \ldots q_n}$ determining the leading dependence on $R$ of $M_{q_1 q_2\ldots q_n}(R)$ is
\begin{equation}\label{e19}
 \Gamma_{q_1q_2\ldots q_n} = 2 x_{q_1q_2\ldots q_n},
\end{equation}
where $x_{q_1 q_2 \ldots q_n}$ is the scaling exponent of the $\sigma$-model operator $(q_1,\ldots, q_n)$ for the correlator (\ref{e5.3}). This is a generalization of Eq.~(\ref{e17}). As with Eq.\ (\ref{e17}), the relation~(\ref{e19}) is expected to be valid only for $q_i$ not too large; probably, the condition is $q_i \le - c_i/2$ for all $i$.

Let us sketch an RG argument in favor of Eq.~(\ref{e19}). We expect that the quantity (\ref{e18}) is represented in field-theory language as a correlation function of two local operators (at points ${\bf r}_1$ and ${\bf r}_2$, respectively), each of which has the same scaling properties as $\nu_1^{q_1} \cdots \nu_n^{q_n}$. Performing an RG transformation that reduces the scale $R$ down to a microscopic scale, we will then get a factor $R^{-2x_{q_1 q_2 \ldots q_n}}$. After this the correlation function becomes of the order of unity; thus, we obtain (\ref{e19}). Possibly, a rigorous proof may be constructed for class A by a generalization of the formula of Ref.~[\onlinecite{klesse01}].

It should be stressed that we do not expect the correlation functions Eq.~(\ref{e18}) to show pure scaling: as we  pointed out, not even the moments of the conductance show it. \cite{janssen99}

\section{Classes with O(1) and U(1) additional degrees of freedom}
\label{s12}

There are five symmetry classes with $\sigma$-model target spaces that either have two connected components and thus an associated $\mathbb{Z}_2 = \text{O}(1)$ degree of freedom [classes D and DIII], or have $\mathbb{Z}$ for their fundamental group due to the presence of a U(1) degree of freedom [classes BDI, CII, AIII]. These degrees of freedom complicate the application of our Weyl-symmetry argument.

We mention in passing that the classes at hand are the five symmetry classes that feature topological insulators in 1D (precisely because, owing to the O(1) and U(1) degrees of freedom, their $\sigma$-model spaces have the said topological properties). Below we briefly outline our present understanding of the Weyl-symmetry issue for these classes and the open questions.

\subsection{Classes D and DIII}\label{s12.1}

The target manifolds of the $\sigma$-models for these symmetry classes consist of two disjoint parts [${\text O}(1) = \mathbb{Z}_2$ degree of freedom]. In general, the $\sigma$-model field can ``jump'' between the two components, thereby creating domain walls. The arguments based on the Weyl symmetry in the form presented above apply directly if such domain walls are prohibited (i.e.\ if the $\sigma$-model field stays within a single component of the manifold). There are several situations when this is the case:
\begin{itemize}
\item
The DMPK model of a quasi-1D wire does not include domain walls. \cite{gruzberg05} This explains the agreement between our symmetry result and the DMPK results for these two classes;
\item
The O(1) version of the Chalker-Coddington network model in 2D; \cite{chalker02}
\item
A good metal in 2D. (In this case domain walls are, strictly speaking, present but their effect is exponentially small and thus expected to be negligible.)
\end{itemize}

Note that the Weyl-group invariance of the LDOS moments for the classes D and DIII yields the symmetry point $q_* / 2 = - c_1 / 2 = 0$. This implies that the distribution function $P(\ln \nu)$ is symmetric under $\ln \nu \to - \ln \nu$ (see Eq. (\ref{e1.4}) with $q_* = 0$), i.e.\ $\ln\nu = 0$ is the most probable (or typical) value. This result is incompatible with exponential localization of the eigenstates, which would imply exponentially small typical LDOS values. We thus arrive at the conclusion that, in the absence of domain walls, systems described by the $\sigma$-model for class D or DIII cannot have a localized phase. The models listed in the previous paragraph exemplify this general statement.

In the case of a good 2D metal in class D or DIII, the scaling behavior can be found by perturbative RG, with the smallness of the inverse conductance $1/g \ll 1$ ensuring the validity of the loop expansion. In particular, the one-loop RG calculation of the average DOS scaling yields \cite{bocquet00} $\langle\nu\rangle \propto \ln L \propto g(L)$. We know that the scaling exponents for the LDOS moments depend quadratically on $q$ in one-loop approximation (which is governed by the quadratic Laplace-Casimir operator). Therefore, in view of the $q \to -q$ Weyl symmetry, we expect the LDOS moments to behave as
\begin{equation}\label{e20}
 \langle\nu^q\rangle \propto (\ln L)^{q^2}.
\end{equation}
It should of course be possible to check this directly by a numerical calculation.

\subsection{Chiral classes}\label{s12.2}

For the chiral classes, the situation is even more subtle. We expect that the Weyl-group invariance should show up most explicitly in operators that are scalars with respect to the additional U(1) degree of freedom. The LDOS moments, however, do not belong to this category. We leave the SUSY-based classification of operators and the investigation of the impact of the Weyl-group invariance to future work.

\vfill\eject

\section{Summary and outlook}\label{s13}

In this paper we have developed a classification of composite operators without spatial derivatives at Anderson-transition critical points in disordered systems. These operators represent observables describing correlations of the local density of states (or wave-function amplitudes). Our classification is motivated by the Iwasawa decomposition for the (complexification of the) supersymmetric $\sigma$-model field. The Iwasawa decomposition has the attractive feature that it gives rise to spherical functions which have the form of ``plane waves'' when expressed in terms of the corresponding radial coordinates. Viewed as composite operators of the $\sigma$-model, these functions exhibit pure-power scaling at criticality. Alternatively, and in fact more appropriately, the same operators can be constructed as highest-weight vectors.

We further showed that a certain Weyl-group invariance (due to the Harish-Chandra isomorphism) leads to numerous exact symmetry relations among the scaling dimensions of the composite operators. Our symmetry relations generalize those derived earlier for the multifractal exponents of the leading operators.

While we focused on the Wigner-Dyson unitary symmetry class (A) in most of the paper, we have also sketched the generalization of our results to some other symmetry classes. More precisely, our results are directly applicable to five (out of the ten) symmetry classes: the three Wigner-Dyson classes (A, AI, AII) and two of the Bogoliubov-de Gennes classes (C and CI). Moreover, they should also be valid for the remaining two Bogoliubov-de Gennes classes (D and DIII), as long as $\sigma$-model domain walls are suppressed (i.e.\ the $\sigma$-model field stays within a single component of the manifold). Our results imply that in this situation the system is protected from Anderson localization. In other words, localization in the symmetry classes D and DIII may take place only due to the appearance of domain walls.

We have further explored the relation of our results for the LDOS (or wave-function) correlators to transport characteristics. We have constructed transport observables that are counterparts of the composite operators for wave-function correlators and conjectured a relation between the scaling exponents.

Our work opens a number of further research directions; here we list some of them.
\begin{enumerate}

\item[(i)]
Verification of our predictions by numerical simulation of systems housing critical points of various dimensionalities, symmetries, and topologies would be highly desirable. While the LDOS multifractal spectra have been studied for a considerable number of critical points, the numerical investigation of the scaling of subleading operators is still in its infancy. Preliminary numerical results for the spectra of scaling exponents of the moments $\langle A_2^q\rangle$ and $\langle A_3^q\rangle$ at the quantum Hall critical point \cite{bera-evers-unpublished} do support our predictions. Furthermore, it would be very interesting to check numerically our predictions for the scaling of transport observables.
\item[(ii)]
As mentioned in Sec.~\ref{s12.2}, it remains to be seen to what extent our results can be generalized to the chiral symmetry classes, and what their implications for observables will be.
\item[(iii)]
In this work, we have studied critical points of non-interacting fermions. In some cases the electron-electron interaction is RG-irrelevant at the fixed point in question, so that the classification remains valid in the presence of the interaction. An example of such a situation is provided by the integer quantum Hall critical point with a short-range electron-electron interaction. \cite{Lee96,Wang00,burmistrov11} However, if the interaction is of long-range (Coulomb) character, the system is driven to another fixed point. (This also happens in the presence of short-range interactions for fixed points with spin-rotation symmetry: in this case, the Hartree-Fock cancelation of the leading term in the two-point function (\ref{e1.5}) does not take place.) The classification of operators and relevant observables at such interacting fixed points, as well as the analysis of possible implications of the Weyl-group invariance, remain challenging problems for future research.
\end{enumerate}

\section{Acknowledgments}

We thank V.~Serganova for useful discussions, and S.~Bera and F.~Evers for informing us of unpublished numerical results. \cite{bera-evers-unpublished} This work was supported by DFG SPP 1459 ``Graphene'' and SPP 1285 ``Semiconductor spintronics'' (ADM). IAG acknowledges the DFG Center for Functional Nanostructures for financial support of his stay in Karlsruhe, and the NSF Grants No. DMR-1105509 and No. DMR-0820054. ADM acknowledges support by the Kavli Institute for Theoretical Physics (University of California, Santa Barbara) during the completion of this work. MRZ acknowledges financial support by the Leibniz program of the DFG.

\appendix

\section{Young diagrams, tableaux, and symmetrizers}
\label{app:notation}

\begin{figure}[t]
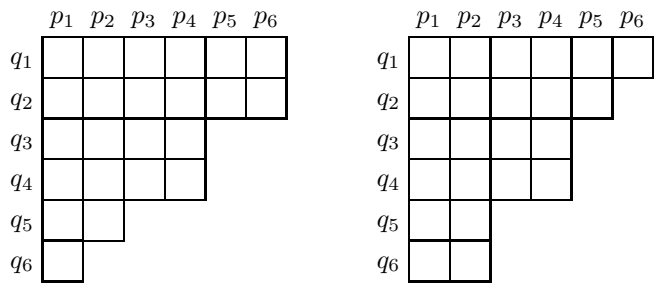

\ytableausetup{mathmode, boxsize=1.5em}
\begin{ytableau}
\none & \none[p_1] & \none[p_2] & \none[p_3] & \none[p_4] & \none[p_5] & \none[p_6] \\
\none[q_1] & & & & & & \\
\none[q_2] & & & & & & \\
\none[q_3] & & & & \\
\none[q_4] & & & & \\
\none[q_5] & & \\
\none[q_6] &
\end{ytableau}
\hfill
\begin{ytableau}
\none & \none[p_1] & \none[p_2] & \none[p_3] & \none[p_4] & \none[p_5] & \none[p_6]\\
\none[q_1] & & & & & &\\
\none[q_2] & & & & &\\
\none[q_3] & & & &\\
\none[q_4] & & & &\\
\none[q_5] & &\\
\none[q_6] & &
\end{ytableau}
\caption{Young diagram $\lambda = (6^2, 4^2, 2, 1) = [6,5,4^2,2^2]$ (left) and its conjugate $\tilde \lambda = (6,5,4^2,2^2) = [6^2, 4^2, 2, 1]$ (right).}
\label{Young-diagrams}
\end{figure}

In this paper we use a standard notation for Young diagrams, see, for example, Ref.\ [\onlinecite{Fulton}]. Thus, the Young diagram corresponding to the partition $q_1 + q_2 + \ldots + q_n$ (where the integers $q_j$ are subject to $q_1 \geq q_2 \geq \ldots \geq q_n \geq 0$) is denoted by $\lambda = (q_1, q_2, \ldots, q_n)$, and consists of left-aligned rows with the top row containing $q_1$ boxes, the next row containing $q_2$ boxes, etc. Another notation that we will use is $\lambda = (l_1^{b_1}, \dots, l_s^{b_s})$ to denote the partition that has $b_i$ copies of the integer $l_i$, $1 \leq i \leq s$. For an example, on the left in Fig.\ \ref{Young-diagrams} we show the Young diagram $\lambda = (6, 6, 4, 4, 2, 1) = (6^2, 4^2, 2, 1)$ with 6 rows and 4 distinct row lengths.

At the top of the diagram in Fig.\ \ref{Young-diagrams} we also display the numbers of boxes $p_1, p_2, \ldots, p_m$ in each column. Like the numbers $q_1, \ldots, q_n$, these completely specify the diagram, and we will use (for the same diagram) the alternative notation $\lambda = [p_1, p_2, \ldots, p_m] = [k_1^{a_1}, \dots, k_s^{a_s}]$, where the second notation means that the partition by the integers $p_i$ has $a_j$ copies of the integer $k_j$, $1 \leq j \leq s$. In this notation the diagram shown in Fig.\ \ref{Young-diagrams} on the left is $\lambda = [6,5,4^2,2^2]$. The numbers $p_i$ also define the conjugate diagram $\tilde \lambda = (p_1, p_2, \ldots, p_m)$. For illustration, the diagram $\tilde \lambda = (6,5,4^2,2^2) = [6^2,4^2, 2, 1]$ is shown in Fig.~\ref{Young-diagrams} on the right. (Notice that the number $s$ of distinct parts is the same for a Young diagram and its conjugate, and is the same as the number of ``corners'' on the boundary of the diagram.) The number of boxes of $\lambda$ and $\tilde\lambda$, called the \emph{size} of $\lambda$, is denoted by
\begin{align}
 |\lambda| = |\tilde \lambda| = \sum_{i=1}^n q_i = \sum_{i=1}^m p_i = N .
\end{align}

For a given Young diagram, the integers $l_i$ and $a_i$ are related by
\begin{align}\label{a-l}
 a_1 &= l_s, \cr
 a_2 &= l_{s-1} - l_s, \cr
 \vdots \cr
 a_s &= l_1 - l_2.
\end{align}
Solving this for $l_i$ gives
\begin{align}\label{l-a}
 l_1 &= a_1 + a_2 + \ldots + a_s, \cr
 l_2 &= a_1 + a_2 + \ldots + a_{s-1}, \cr
 \vdots \cr
 l_s &= a_1.
\end{align}
Similar relations exist between $k_i$ and $b_i$:
\begin{align}
 b_1 &= k_s, \cr
 b_2 &= k_{s-1} - k_s, \cr
 \vdots \cr
 b_s &= k_1 - k_2. \label{b-k} \\
 k_1 &= b_1 + b_2 + \ldots + b_s, \cr
 k_2 &= b_1 + b_2 + \ldots + b_{s-1}, \cr
 \vdots \cr
 k_s &= b_1. \label{k-b}
\end{align}

Young diagrams are used to label irreducible representations (irreps) of the permutation groups and some classical matrix groups.
%Initially, we only deal with symmetry class A, where the symmetry group %is $G = \mathrm{U}(n)$. In that case, a Young diagram $\lambda$ ...
%an irrep of U$(n)$ with an arbitrary $n$.
Irreps of $G = \mathrm{U}(n)$ of polynomial type are in one-to-one correspondence with Young diagrams that have at most $n$ rows. The eigenvalue of the quadratic Casimir operator in the irrep of $\mathrm{U}(n)$ with Young diagram $\lambda = (q_1, q_2, \ldots, q_n)$ is
\begin{align}\label{a2}
 a_2(\lambda, n) = \sum_{i=1}^n q_i (q_i + n + 1 - 2i).
\end{align}
It is known that the quadratic Casimir eigenvalue for the conjugate Young diagram $\tilde \lambda$ is related to the one for $\lambda$ by
\begin{align}
 a_2(\tilde\lambda, n) = -a_2(\lambda, -n).
\end{align}

Next we need the notion of Young tableaux. A tableau $T$ is a Young diagram $\lambda$ with each of its boxes filled with a positive integer from the set $\{1, 2, \ldots, N = |\lambda| \}$. A tableau is called semistandard if the integers in the boxes \textit{i}) weakly increase from left to right along each row and \textit{ii}) strictly increase from top to bottom along each column. The minimal semistandard tableau (denoted by $T_{\text{min}}$) for a given shape $\lambda = [k_1^{a_1},\dots,k_s^{a_s} ]$ is the one where all the integers in the first row are 1, in the second row 2, and so on, up to $k_1$ in the last row. A semistandard tableau is called standard if it is filled according to the above rules so that each number from the set $\{ 1, \ldots, N \}$ occurs exactly once. A normal tableau (which we denote by $T_0$) is a standard Young tableau in which the numbers are in order, left to right and top to bottom. If a tableau $T$ is obtained by filling a Young diagram $\lambda$, we say that $\lambda$ is
the shape of $T$. To give an example, Figure \ref{Young-tableaus} shows a semistandard, the minimal, a standard, and the normal tableau of shape $(6, 4^2, 2)$.

\begin{figure}[t]
\ytableausetup{mathmode, boxsize=1.25em}
\subfloat[Semistandard tableau]
{\label{semistandard}
\ytableaushort[]{123447,2355,3467,56}}
\hskip 15mm
\subfloat[Minimal tableau]
{\label{minimal}
\ytableaushort[]{111111,2222,3333,44}} \\
\subfloat[Standard tableau]
{\label{standard}
\ytableaushort[]{137{12}{13}{15},25{10}{14},48{11}{16},69}}
\hskip 15mm
\subfloat[Normal tableau]
{\label{normal}
\ytableaushort[]{123456,789{10},{11}{12}{13}{14},{15}{16}}}
\caption{Young tableaux}
\label{Young-tableaus}
\end{figure}

The permutation group $S_N$ acts on %{\bf standard}
tableaux with $N$ boxes by permuting the integers in the boxes. If $\sigma \in S_N$, we denote by $\sigma T$ the tableau which has the number $\sigma(i)$ in the box where $T$ has $i$. For each Young diagram $\lambda$ of size $N$ we define $R(\lambda)$ and $C(\lambda)$ as the subgroups of $S_N$ that preserve the rows and columns of $\lambda$, respectively. One can consider formal linear combinations of the elements of $S_N$ (these form what is known as the group algebra ${\cal A}(S_N)$ of $S_N$) and define the following operators:
\begin{align}
 a_\lambda &= \sum_{\sigma \in R(\lambda)} \sigma, & b_\lambda = \sum_{\tau \in C(\lambda)} \text{sgn}(\tau) \tau .
\end{align}
When acting on a tableau $T$, the operator $a_\lambda$ symmetrizes all the numbers in $T$ along its rows. Similarly, $b_\lambda$ antisymmetrizes entries of a tableau along the columns. Finally, the Young symmetrizers are defined as the products
\begin{align}\label{Young-symmetrizer}
 c_\lambda = b_\lambda a_\lambda .
\end{align}
Sometimes one uses an alternative definition of the Young symmetrizers where the order of the operations of symmetrization and antisymmetrization is reversed:
\begin{align}\label{Young-symmetrizer-alternative}
 {\tilde c}_\lambda = a_\lambda b_\lambda.
\end{align}

All operators $a_\lambda$, $b_\lambda$, $c_\lambda$, and ${\tilde c}_\lambda$ are idempotent; this means that their squares are proportional to the operators themselves:
\begin{align}\label{idempotent}
 a_\lambda^2 &= n_R a_\lambda, & b_\lambda^2 &= n_C b_\lambda, \cr
 c_\lambda^2 &= n_\lambda c_\lambda, & {\tilde c}_\lambda^2 &= n_\lambda {\tilde c}_\lambda ,
\end{align}
where $n_R$ and $n_C$ are the orders of the subgroups $R(\lambda)$ and $C(\lambda)$, and $n_\lambda$ is another positive integer.

For Young diagrams of type $(q) = [1^q]$ the Young symmetrizer reduces to the total symmetrizer along the single row. Similarly, for the diagrams of type $(1^p) = [p]$ the Young symmetrizer is the total antisymmetrizer along the single column:
\begin{align}
 c_{(q)} &= a_{(q)}, && c_{[p]} = b_{[p]}.
\end{align}

To illustrate these operators, consider the normal tableau $T_0$ for $\lambda = (2,1)$. In that case,
\begin{align}
 \ytableausetup{boxsize=1.25em}
 a_{(2,1)} \ytableaushort{12,3} &= \ytableaushort{12,3} + \ytableaushort{21,3} \\
 b_{(2,1)} \ytableaushort{12,3} &= \ytableaushort{12,3} - \ytableaushort{32,1} \\
 \label{YS-example}
 c_{(2,1)} \ytableaushort{12,3} &= b_{(2,1)} \ytableaushort{12,3} +
 b_{(2,1)} \ytableaushort{21,3} \nonumber \\
 &= \ytableaushort{12,3} - \ytableaushort{32,1} + \ytableaushort{21,3} - \ytableaushort{31,2} \\
 {\tilde c}_{(2,1)} \ytableaushort{12,3} &= a_{(2,1)} \ytableaushort{12,3} - a_{(2,1)} \ytableaushort{32,1} \nonumber \\
 &= \ytableaushort{12,3} + \ytableaushort{21,3} - \ytableaushort{32,1} - \ytableaushort{23,1}
\end{align}

For the purposes of this paper we also need to consider tableaux filled by points and wave-function symbols rather than integers. The action of the permutation groups and Young symmetrizers on such tableaux is defined in the same way as on tableaux filled by integers: the points and wave functions are permuted according to their positions in a tableau. Now, suppose we have a young diagram $\lambda$ and two tableaux, $T_\psi$ and $T_r$ of shape $\lambda$, filled with wave functions and points, respectively. We can define a pairing of these two tableaux as the following product of wave functions:
\begin{align}\label{pairing}
 \Psi_\lambda(T_\psi, T_r) = \prod_{i \in \lambda} \psi_i({\bf r}_i) ,
\end{align}
where $i$ runs over the boxes of the diagram $\lambda$. The tableaux used in this definition need not be standard or semistandard but can be arbitrary. For example, for the following two tableaux
\begin{align}
 \ytableausetup{boxsize=1.25em}
 T_\psi &= \ytableaushort[\psi_]{245,15,3} && T_r =  \ytableaushort[{\bf r}_]{123,13,4} \;,
\end{align}
the corresponding product of wave functions is
\begin{align*}
 \Psi_{(3,2,1)}(T_\psi, T_r) = \psi_1({\bf r}_1) \psi_2({\bf r}_1) \psi_3({\bf r}_4) \psi_4({\bf r}_2) \psi_5^2({\bf r}_3) .
\end{align*}

When an element $s \in {\cal A}(S_{|\lambda|})$ of the group algebra of the symmetric group $S_{|\lambda|}$ acts on one of the arguments of $\Psi_\lambda(T_\psi, T_r)$, we understand, say
\begin{align}\label{group-algebra-action}
 \Psi_\lambda(T_\psi, s T_r) ,
\end{align}
as the linear combination of the corresponding products. It is easy to derive some useful properties of such actions. First of all, it is clear that if we permute the entries in both tableaux $T_\psi$ and $T_r$ in the same way, then we do not change the pairing of these two tableaux:
\begin{align}
 \Psi_\lambda(\sigma T_\psi, \sigma T_r) = \Psi_\lambda(T_\psi, T_r), \quad \sigma \in S_{|\lambda|}.
\end{align}
Now take the inverse $\sigma^{-1}$ of a permutation $\sigma \in S_{|\lambda|}$, and apply it to both tableaux in $\Psi_\lambda(T_\psi, \sigma T_r)$. Then
\begin{align}\label{permutation}
 \Psi_\lambda(T_\psi, \sigma T_r) = \Psi_\lambda(\sigma^{-1} T_\psi, T_r).
\end{align}
Next, if the permutation $\sigma$ in the last equation runs over either of the subgroups $R(\lambda)$ or $C(\lambda)$, its inverse $\sigma^{-1}$ does the same. Moreover, the parities of $\sigma$ and $\sigma^{-1}$ are the same. Therefore, summing Eq.\ (\ref{permutation}) over $R(\lambda)$ or over $C(\lambda)$ with appropriate sign factors, we get
\begin{align}
 \Psi_\lambda(T_\psi, a_\lambda T_r) &= \Psi_\lambda(a_\lambda T_\psi, T_r), \\
 \Psi_\lambda(T_\psi, b_\lambda T_r) &= \Psi_\lambda(b_\lambda T_\psi, T_r).
\end{align}
Finally, using the last two equations it is easy to obtain
\begin{align}
 \Psi_\lambda(T_\psi, c_\lambda T_r) &= \Psi_\lambda(b_\lambda T_\psi, a_\lambda T_r) = \Psi_\lambda({\tilde c}_\lambda T_\psi, T_r), \label{YS-action-1} \\
 \Psi_\lambda(T_\psi, {\tilde c}_\lambda T_r) &= \Psi_\lambda(a_\lambda T_\psi, b_\lambda T_r) = \Psi_\lambda(c_\lambda T_\psi, T_r).
 \label{YS-action-2}
\end{align}

The combinations of wave functions that play a special role in the paper are $\Psi_\lambda(T, c_\lambda T)$, where $T$ is a standard tableau of shape $\lambda$. It is these particular combinations that lead to pure scaling operators in the $\sigma$-model, see Section \ref{s6}. For example, we can take both $T_\psi$ and $T_r$ to be the normal tableau and obtain $\Psi_\lambda(T_0, c_\lambda T_0)$.

Let us look at a few simple examples. If we take both $T_\psi$ and $T_r$ to be the normal tableau for the diagram $(2,1)$ and act on $T_r$ by the Young symmetrizer $c_{(2,1)}$ of (\ref{Young-symmetrizer}), then by using (\ref{YS-example}) we get
\begin{align}
 &\Psi_{(2,1)}(T_0, c_{(2,1)}T_0) = \psi_1({\bf r}_1) \psi_2({\bf r}_2) \psi_3({\bf r}_3) \cr
 &- \psi_1({\bf r}_3) \psi_2({\bf r}_2) \psi_3({\bf r}_1) + \psi_1({\bf r}_2) \psi_2({\bf r}_1) \psi_3({\bf r}_3) \cr
 &- \psi_1({\bf r}_3) \psi_2({\bf r}_1) \psi_3({\bf r}_2) \cr
 &= \psi_2({\bf r}_2) D_2({\bf r}_1, {\bf r}_3) + \psi_2({\bf r}_1) D_2({\bf r}_2, {\bf r}_3).
\end{align}
It is an easy exercise to show that one has the same expression for $\Psi_{(2,1)}({\tilde c}_{(2,1)}T_0, T_0)$.

In the notation of Eq.\ (\ref{group-algebra-action}) the Slater determinants (\ref{e4.14}) can be written as
\begin{align}
 D_p({\bf r}_1,\ldots, {\bf r}_p) = \Psi_{(1^p)}(T_0, b_{(1^p)}T_0).
\end{align}

If we build our product of wave functions by taking the minimal semistandard tableau of a given shape $\lambda = [k_1^{a_1}, \dots, k_s^{a_s}]$ for both $T_\psi$ and $T_r$, then the operation of symmetrization along the rows is clearly redundant, and we get
\begin{align}\label{Psi-minimal}
 &\Psi_\lambda(T_{\text{min}}, c_\lambda T_{\text{min}}) \propto
 \Psi_\lambda(T_{\text{min}}, b_\lambda T_{\text{min}}) \cr
 &\quad = \Psi_\lambda(b_\lambda T_{\text{min}}, T_{\text{min}}) = D_{k_1}^{a_1} \cdots D_{k_s}^{a_s},
\end{align}
where each determinant $D_j$ is evaluated on wave functions $\psi_1, \ldots, \psi_j$ at the points ${\bf r}_1, \ldots, {\bf r}_j$. Adopting the notation $\lambda = (q_1, q_2, \ldots, q_n)$, an alternative form of this expression is
\begin{align}\label{Psi-minimal-1}
 &\Psi_\lambda(T_{\text{min}}, c_\lambda T_{\text{min}}) \propto D_1^{q_1 -q _2} D_2^{q_2 - q_3} \cdots D_n^{q_n}.
\end{align}

\section{Construction of highest-weight vectors}
\label{appendix-hwv}

In this appendix we construct the scaling operators (\ref{e6.8}), (\ref{e6.25}) by using the idea of the highest-weight vector sketched in Sec.\ \ref{s8}. As was discussed there, we first focus on linear functions $\mu_Y(Q)$ (\ref{e8.7}) of the matrix elements of the $\sigma$-model field $Q$ specified by a matrix $Y$: $\mu_Y(Q) = \text{Tr} (YQ)$. Let us denote by $E_{ij}$ the matrix which contains the number one at the intersection of the $i$-th row with the $j$-th column and zeros everywhere else. Such matrices are sometimes called ``matrix units''. Individual matrix elements of $Q$ can be written as
\begin{align}\label{eB.1}
 \mu_{E_{ij}}(Q) = Q_{ji} .
\end{align}

We begin with two simple examples: functions on a sphere $S^2$ and on a hyperboloid $H^2$; these are symmetric spaces of compact and non-compact type, respectively.

\subsection{Functions on a sphere}\label{subsec:sphere}

Consider the space of functions on the two-sphere
\begin{align}
 G/K = \text{U}(2) / \text{U}(1) \times \text{U}(1) = \text{SU}(2)/\text{U}(1) = S^2 .
\end{align}
To make the presentation here similar to the general case considered later, we represent points on the sphere by the matrix
\begin{align}
 Q = g \Lambda g^{-1},
\end{align}
where $g \in \text{SU}(2)$ and $\Lambda = \sigma_3$ is the third Pauli matrix. Using a parametrization of SU$(2)$ by Euler angles,
\begin{align}
 g = \begin{pmatrix*}[r]
 e^{-i(\phi + \psi)/2} \cos \frac{\theta}{2} & - e^{-i(\phi - \psi)/2} \sin \frac{\theta}{2} \\
 e^{i(\phi - \psi)/2} \sin \frac{\theta}{2} & e^{i(\phi + \psi)/2} \cos \frac{\theta}{2}
\end{pmatrix*}
\end{align}
we get
\begin{align}
 Q = \begin{pmatrix*} x_3 & x_1 - i x_2 \\ x_1 + i x_2 & - x_3
\end{pmatrix*},
\end{align}
where
\begin{align}
 x_1 &= \sin\theta \cos\phi, & x_2 &= \sin\theta \sin\phi, & x_3 &= \cos\theta
\end{align}
are the three basic functions which arise by restricting the Cartesian coordinates of the Euclidean space $\mathbb{R}^3$ to the sphere $x_1^2 + x_2^2 + x_3^2 = 1$.

Let us choose $X_k = \sigma_k/(2i)$ as our system of generators of the Lie algebra su(2). The standard choice for the Cartan generator is $X_3$, and the raising and lowering generators (in the complexification sl$(2, \mathbb{C})$ of su$(2)$) are $X_\pm = i X_1 \mp X_2$. Notice that
\begin{align}
 X_+ = E_{12} , \quad X_- = E_{21} .
\end{align}
Now the function
\begin{align}
 \varphi_1(Q) = \mu_{X_+}(Q) = Q_{21} = x_1 + i x_2 = \sin\theta \, e^{i\phi}
\label{eB.8}
\end{align}
is a highest-weight vector for the SU$(2)$ action. Indeed, by Eq.\ (\ref{e8.8}) we have
\begin{align}
 \widehat{X}_+ \varphi_1 &= \mu_{[X_+, X_+]} = 0, \\ \widehat{X}_3 \varphi_1 &= \mu_{[X_3, X_+]} = \mu_{-iX_+} = -i\varphi_1 .
\end{align}
Powers of this function,
\begin{align}
 \varphi_l(Q) \equiv \varphi_1^l(Q) = (x_1 + i x_2)^l = \sin^l \theta \, e^{il\phi} ,
\end{align}
are also highest-weight vectors. To make them globally well defined on the sphere, the power $l$ has to be a non-negative integer. Constant multiples of $\varphi_l$ are known as the spherical harmonics $Y_{ll}$ in the irreducible representation of SU$(2)$ of dimension $2l + 1$.

There exist other choices of Cartan and nilpotent subalgebras. For example, if we chose $X_1$ as the Cartan generator, and $X'_+ = -X_3 + i X_2$ as the raising operator, then the (linear) highest-weight vector would be
\begin{align}\label{eB.12}
 \varphi_1' &= \frac{i}{2}(Q_{11} - Q_{21} + Q_{12} - Q_{22}) \cr
 & = i(x_3 + i x_2) = i\cos\theta - \sin\theta \sin\phi.
\end{align}
Similarly, the choice of $X_2$ as the Cartan generator would lead to the highest-weight vector
\begin{align}\label{eB.13}
 \varphi_1'' = x_3 + i x_1 = \cos\theta + i \sin\theta \cos\phi.
\end{align}
Both $\varphi_1'$ and $\varphi_1''$ can be raised to non-negative integer powers to produce other highest-weight vectors.

In this example of functions on a compact symmetric space, all three choices of Cartan subalgebra or highest-weight vector are equivalent and can be transformed into each other by an element of $G = \mathrm{SU}(2)$. (In fact they are just three ``points'' on an $\mathrm{SU}(2)$-orbit of Cartan subalgebras or highest-weight vectors.) This will not be the case in our next example of functions on the two-hyperboloid.

\subsection{Functions on a hyperboloid}\label{subsec:hyperboloid}

We now consider the space of functions on a non-compact analog of the two-sphere, the two-hyperboloid
\begin{align}
 G/K = \text{SU}(1,1)/\text{U}(1) = H^2.
\end{align}
One may view this space as a non-compact variant of the sphere $S^2$, by analytically continuing the compact angle $\theta$ to the non-compact radial variable on $H^2$ (denoted by the same symbol $\theta$). If we make the replacement $\theta \to i\theta$ in the function (\ref{eB.8}), we get $i \sinh \theta \, e^{i\phi}$. While this function is a highest-weight vector for some choice of the Cartan subalgebra, it is not positive on the hyperboloid, so it cannot be raised to an arbitrary complex power. However, there exist other, inequivalent choices of the Cartan subalgebra which do give the desired positivity property. In fact, if we analytically continue $\theta \to i\theta$ in, say, Eq.\ (\ref{eB.13}), we get the highest-weight vector $\cosh\theta - \sinh\theta \cos\phi$, which is strictly positive on $H^2$ and, therefore, can be raised to an arbitrary complex power. Here is how it is done more formally.

Matrices $g \in \text{SU}(1,1)$ satisfy the relation
\begin{align}\label{g-inverse}
 \quad g^{-1} = \sigma_3 g^\dagger \sigma_3,
\end{align}
and can be parametrized in terms of generalized Euler angles as
\begin{align}\label{Euler-3}
 g &= \begin{pmatrix*}[r]
 e^{i(\phi + \psi)/2} \cosh \frac{\theta}{2} & -i e^{i(\phi - \psi)/2} \sinh \frac{\theta}{2} \\
 i e^{-i(\phi - \psi)/2} \sinh \frac{\theta}{2} & e^{-i(\phi + \psi)/2} \cosh \frac{\theta}{2}
\end{pmatrix*}.
\end{align}
Elements of the coset space $G/K = H^2$ are represented by matrices
\begin{align}
 Q &= g \sigma_3 g^{-1} = \begin{pmatrix*} x_3 & i x_1 + x_2 \\ i x_1 - x_2 & - x_3 \end{pmatrix*} .
\end{align}
The matrix elements $x_1, x_2, x_3$ may be viewed as the Cartesian coordinates of the Euclidean space $\mathbb{R}^3$ restricted to the hyperboloid $x_3^2 - x_1^2 - x_2^2 = 1$. By adopting the parametrization (\ref{Euler-3}) we express them as
\begin{align}
 x_1 &= \sinh\theta \cos\phi, & x_2 &= \sinh\theta \sin\phi, & x_3 &= \cosh\theta .
\end{align}

The Lie algebra $\mathrm{su}(1,1) \simeq \mathbb{R}^3$ is spanned by the matrices $i X_1$, $i X_2$, and $X_3$. Choosing $X_3$ for the Cartan generator and $X_\pm = i X_1 \mp X_2$ for the nilpotent generators, we get the highest-weight vector
\begin{align}
 \varphi_1 = Q_{21} = i x_1 - x_2 = i \sinh\theta \, e^{i\phi}.
\end{align}
This is the analog of (\ref{eB.8}) for the hyperboloid, and it is not a positive function. To obtain a positive highest-weight vector, we need to choose a linear combination of $i X_1$ and $i X_2$ for the Cartan generator.

Thus let the Cartan generator be $i X_1 \cos\alpha - i X_2 \sin\alpha$ for some choice of parameter $\alpha$. Taking
\begin{align*}
 X_+' &= X_2 \cos\alpha + X_1 \sin\alpha + i X_3
\end{align*}
for the raising operator, we have the following expression for the corresponding highest-weight vector:
\begin{align}
 \varphi_1' &= \mu_{X_+'} = {\textstyle{\frac{1}{2}}}(Q_{11} - Q_{22} + e^{-i\alpha} Q_{12} - e^{i\alpha} Q_{21}) \cr &= x_3 + x_2 \cos\alpha + x_1 \sin\alpha \cr
 &= \cosh\theta + \sinh\theta \sin (\phi + \alpha) ,
\end{align}
which already arose in the closely related context of Eq.\ (\ref{e4.9}). An arbitrary complex power of this positive function is also a highest-weight vector:
\begin{align}
 \varphi_q' = \big( \cosh\theta + \sinh\theta \sin(\phi + \alpha) \big)^q, \quad q \in \mathbb{C}.
\end{align}
and this function (or rather, its extension to the SUSY setting) is the $\sigma$-model scaling operator for the $q$-th power of the local density of states.

While it is clear by inspection that the function $\varphi_1'$ is positive, a more formal proof that generalizes to cases of higher rank is as follows. We write
\begin{align}
 \varphi_1' &= {\textstyle{\frac{1}{2}}} \text{Tr}[(\sigma_3 - i \sigma_2 \cos\alpha - i \sigma_1 \sin\alpha)\, g \sigma_3 g^{-1}] \cr &= {\textstyle{\frac{1}{2}}} \text{Tr}[(1 - \sigma_1 \cos\alpha + \sigma_2 \sin\alpha) g g^\dagger] ,
\end{align}
where we have used the SU$(1,1)$ defining relation (\ref{g-inverse}). The matrix $\Pi = (1 - \sigma_1 \cos\alpha + \sigma_2 \sin\alpha)/2$ is a projection operator: $\Pi^\dagger = \Pi = \Pi^2$. We thus see that the function $\varphi_1'$ is the manifestly positive expectation value of $g g^\dagger > 0$ in the eigenvector of the projector $\Pi$ with eigenvalue 1.

\subsection{Arbitrary $n$, compact case}\label{subsec:compact}

We now come back to the general case of functions on the compact symmetric space for class A:
\begin{align}
 G/K = \text{U}(2n)/ \text{U}(n) \times \text{U}(n) ,
\end{align}
which arises from the use of fermionic replicas. Elements of this coset space, or points on the manifold, are represented by matrices $Q = g \Lambda g^{-1}$, where now
\begin{align}\label{Lambda}
 \Lambda = \Sigma_3 = \begin{pmatrix} \one_n & 0 \\ 0 & -\one_n \end{pmatrix}.
\end{align}

We begin with a choice of root-space decomposition $\mfg = \mfn_+ \oplus \mfh \oplus \mfn_-$, see Eq.\ (\ref{e8.2}). We take $\mfh$ to be spanned by the diagonal matrices and $\mfn_+$ ($\mfn_-$) be spanned by the upper (respectively, lower) triangular matrices in $\mfg = \text{gl}(2n, \mathbb{C})$. Schematically,
\begin{align}
 \mfh &= \begin{pmatrix} * &0 &\ldots &0 &0 \\ 0 &* &\ldots &0 &0 \\
 \vdots &\vdots &\ddots &\vdots &\vdots \\ 0 &0 &\ldots &* &0 \\
 0 &0 &\ldots &0 &* \end{pmatrix}, &
 \mfn_+ = \begin{pmatrix} 0 &* &\ldots &* &* \\ 0 &0 &\ldots &* &* \\
 \vdots & \vdots & \ddots & \vdots & \vdots \\0 & 0 & \ldots & 0 & * \\
 0 & 0 & \ldots & 0 & 0 \end{pmatrix} .
\end{align}
We also need the (refined) Cartan decomposition $\mfg = \mfp_+ \oplus \mfk \oplus \mfp_-$, where $\mfk = \text{gl}(n,\mathbb{C}) \oplus \text{gl}(n, \mathbb{C})$ is the complexified Lie algebra of $K = \text{U}(n) \times \text{U}(n)$, while $\mfp_\pm$ are the eigenspaces of the adjoint (or commutator) action of $\Sigma_3$:
\begin{align}
 \mfk &= \begin{pmatrix} * & \ldots & * & 0 & \ldots & 0 \\
 \vdots & \ddots & \vdots & \vdots & \ddots & \vdots  \\
 * & \ldots & * & 0 & \ldots & 0 \\
 0 & \ldots & 0 & * & \ldots & * \\
 \vdots & \ddots & \vdots & \vdots & \ddots & \vdots  \\
 0 & \ldots & 0 & * & \ldots & *
 \end{pmatrix}, &
 \mfp_+ = \begin{pmatrix} 0 & \ldots & 0 & * & \ldots & * \\
 \vdots & \ddots & \vdots & \vdots & \ddots & \vdots  \\
 0 & \ldots & 0 & * & \ldots & * \\
 0 & \ldots & 0 & 0 & \ldots & 0 \\
 \vdots & \ddots & \vdots & \vdots & \ddots & \vdots  \\
 0 & \ldots & 0 & 0 & \ldots & 0
\end{pmatrix} .
\end{align}
Note the commutation relations
\begin{align}\label{comm-rel-Cartan}
 [\mfk, \mfp_+] &= \mfp_+, & [\mfp_+, \mfp_+] &= 0 ,
\end{align}
as well as the following decomposition of the space $\mfn_+$ of raising operators:
\begin{align}\label{n+decomposition}
 \mfn_+ = \mfp_+ \oplus (\mfn_+ \cap \mfk) .
\end{align}

Our attention now focuses on the space of complex-valued functions $\mu_Y$ (see Eq.\ (\ref{e8.7})) for $Y \in \mfp_+$. We will use such functions as building blocks to construct functions that have the properties of a highest-weight vector, see (\ref{e8.3}). In fact, by the second set of commutation relations in (\ref{comm-rel-Cartan}) any function $\mu_Y$ for $Y \in \mfp_+$ is already annihilated by all first-order differential operators that represent generators of $\mfp_+$:
\begin{align}
 \widehat{X} \mu_Y = \mu_{[X,Y]} = 0 \quad \text{ for } X, Y \in \mfp_+.
\end{align}
However, $\mu_Y$ for general $Y \in \mfp_+$ is not annihilated by all raising operators from $\mfn_+ \cap \mfk$. To implement this annihilation condition, we construct certain polynomials of the matrix elements of $Q$ in the following way.

For $1 \leq i, j \leq n$ we introduce the functions
\begin{align}
 \nu_{ij} = \mu_{E_{i, 2n+1 - j}} = Q_{2n+1 - j, i}.
\end{align}
Notice that the matrix $E_{i, 2n+1 - j} \in \mfp_+$, so the functions $\nu_{ij}$ are exactly of the type discussed in the previous paragraph. Now we will demonstrate that for any integer $m$ in the range $1 \leq m \leq n$ the $m\times m$ determinant
\begin{align}
 f_m = \text{Det} \begin{pmatrix}
 \nu_{11} & \ldots & \nu_{1m} \\ \vdots & \ddots & \vdots \\ \nu_{m1} & \ldots & \nu_{mm} \end{pmatrix}
\end{align}
is a highest-weight vector for the decomposition (\ref{e8.2}).

We first establish that $\widehat{X} f_m = 0$ for all $X \in \mfn_+$. Due to the decomposition (\ref{n+decomposition}) there are two cases to consider. First, let $X \in \mfp_+$. Then, as we have already mentioned, $\widehat{X} \nu_{ij} = 0$ and, therefore, $\widehat{X} f_m = 0$. Now let $X$ be in the space $\mfn_+ \cap \mfk$, which is spanned by $E_{ii'}$ for $1\leq i < i' \leq n$ and $E_{jj'}$ for $n+1 \leq j < j' \leq 2n$. If $X = E_{ii'}$ with $m < i' \leq n$, then we still have $\widehat{X} f_m = 0$, since for all $1 \leq i'',j \leq m$ the commutator
\begin{align}
 [E_{ii'}, E_{i'',2n+1 - j}] = E_{ii'} E_{i'',2n+1 - j}
\end{align}
vanishes due to $i' \not= i''$. Now let $X = E_{ii'}$ with $1 \leq i < i' \leq m$. In this case we obtain
\begin{eqnarray}
 \widehat{E}_{ii'} f_m &=& \frac{d}{dt}\Big|_{t=0} \text{Det} \big[ \text{Tr} \big(e^{t E_{ii'}} E_{i'',2n+1-j} e^{-t E_{ii'}} Q\big)\big]_{i'',j=1}^m \cr &=& \frac{d}{dt}\Big|_{t=0} \text{Det} \big[ \nu_{i''j} + t \delta_{i'i''} \nu_{ij} \big]_{i'',j = 1}^m .
\end{eqnarray}
The matrix under the determinant sign in the last equation factorizes as $(1 + t E_{i'i})\nu$, therefore
\begin{align}
 \widehat{E}_{ii'} f_m &= f_m \frac{d}{dt}\Big|_{t=0} \text{Det} (1 + t E_{i'i}) = 0,
\end{align}
since the determinant in the last equation does not depend on $t$.

It remains to show that $f_m$ is an eigenfunction of the operators from $\mfh$. To this end we express $H \in \mfh$ as $H = \sum_{i=1}^{2n} h_i E_{ii}$, where $h_i \in \mathbb{C}$. This is a diagonal matrix, and so is $e^{tH} = \sum_{i=1}^{2n} e^{t h_i} E_{ii}$. Then we have
\begin{eqnarray}
 \widehat{H} \nu_{ij} &=& \frac{d}{dt}\Big|_{t=0} \sum_{kl} e^{t(h_k - h_l)} \text{Tr} (E_{kk} E_{i, 2n+1-j} E_{ll} Q) \cr
 &=& \frac{d}{dt}\Big|_{t=0}   e^{t(h_i - h_{2n+1-j})} \nu_{ij},
\end{eqnarray}
and, indeed, the property (\ref{e8.3}) follows:
\begin{align}
 \widehat{H} f_m &=  \frac{d}{dt}\Big|_{t=0} \text{Det} \big[e^{t(h_i - h_{2n+1-j})} \nu_{ij} \big]_{i,j = 1}^m \cr
 &= f_m \frac{d}{dt}\Big|_{t=0} e^{t\sum_{i=1}^m (h_i - h_{2n+1-i})} = \lambda_m(H) f_m, \cr \lambda_m(H) &= \sum_{i=1}^m (h_i - h_{2n+1-i}).
\end{align}

Since all functions $f_m$ for $1 \leq m \leq n$ have the highest-weight property, so does the product
\begin{align}
 \varphi_{(q_1,\ldots,q_n)} = f_1^{q_1 - q_2} f_2^{q_2 - q_3} \cdots f_{n-1}^{q_{n-1} - q_n} f_n^{q_n}
\end{align}
for a weakly decreasing sequence of $n$ integers $q_1 \geq q_2 \geq \ldots \geq q_n \geq 0$. The powers in this expression are restricted to be non-negative integers, since the functions $f_m$ are complex-valued. The functions $\varphi_{(q_1,\ldots,q_n)}$ are the most general highest-weight vectors in the present situation.

Note that if $H = \sum_{i=1}^n h_i E_{ii}$ is a diagonal generator of GL$(n,\mathbb{C})$, then
\begin{align}
 \widehat{H} \varphi_{(q_1,\ldots,q_n)} = \Big(\sum_{i=1}^n q_i h_i \Big) \varphi_{(q_1,\ldots,q_n)}.
\end{align}
By standard facts of representation theory it follows that $(q_1,\ldots,q_n)$ my be interpreted as the sequence of numbers determining the Young diagram of an irreducible representation of GL$(n,\mathbb{C})$.

\subsection{Arbitrary $n$, non-compact case}\label{subsec:noncompact}

We now turn to the general non-compact situation for class A and consider the space of functions on
\begin{align}
 G/K = \text{U}(n,n) / \text{U}(n) \times \text{U}(n) ,
\end{align}
which results from the use of bosonic replicas. Points on this manifold are represented by matrices $Q = g \Lambda g^{-1}$, where $\Lambda = \Sigma_3$ is defined in Eq.\ (\ref{Lambda}). Elements of the pseudo-unitary group $\text{U}(n,n)$ satisfy
\begin{align}\label{g-inverse-general-n}
 g^{-1} = \Sigma_3 g^\dagger \Sigma_3 .
\end{align}
Hence the functions $\mu_Y(Q)$ can be rewritten as
\begin{align}\label{mu-Y}
 \mu_Y(Q) &= \text{Tr}(Y g \Sigma_3 g^{-1}) = \text{Tr}(g g^\dagger \Sigma_3 Y).
\end{align}

We will try to follow the development of the compact case as much as possible. One major change comes from the fact that by the hyperbolic nature of the Lie algebra $\mfg = \text{u}(n,n)$, there exist several $G$-inequivalent choices of Cartan subalgebra $\mfh$. For our purposes, the good choice to consider is as follows.

We make an orthogonal transformation of the standard basis $\{ e_i \}$ of $\mathbb{C}^{2n}$ to introduce a new basis $\{\tilde{e}_i\}$:
\begin{align}
 \tilde{e}_j &= \frac{e_j + e_{j+n}}{\sqrt{2}}, &
 \tilde{e}_{2n+1-j} = \frac{e_j - e_{j+n}}{\sqrt{2}} ,
\end{align}
where $1 \leq j \leq n$. We then define linear operators $\tilde{E}_{ij}$ on $\mathbb{C}^{2n}$ by the relation $\tilde{E}_{ij} \tilde{e}_k = \delta_{jk} \tilde{e}_i$. Thus, in the new basis the operators $\tilde{E}_{ij}$ are the ``matrix units''. They can be expressed in terms of the matrix units with respect to the original basis:
\begin{align*}
 \tilde{E}_{ij} &= {\textstyle{\frac{1}{2}}} (E_{ij} + E_{i + n, j} + E_{i, j + n} + E_{i + n, j + n}), \cr
 \tilde{E}_{2n+1-i,j} &= {\textstyle{\frac{1}{2}}} (E_{ij} - E_{i + n, j} + E_{i, j + n} - E_{i + n, j + n}), \cr
 \tilde{E}_{i,2n+1-j} &= {\textstyle{\frac{1}{2}}}(E_{ij} + E_{i + n, j} - E_{i, j + n} - E_{i + n, j + n}), \cr
 \tilde{E}_{2n+1-i,2n+1-j} &= {\textstyle{\frac{1}{2}}}(E_{ij} - E_{i + n, j} - E_{i, j + n} + E_{i + n, j + n}).
\end{align*}

With these conventions, let us choose the Cartan subalgebra $\mfh$ and the subalgebra $\mfn_+$ of raising operators as follows:
\begin{align}
 \mfh &= \text{span}_{\mathbb C}\big\{\tilde{E}_{jj}, 1 \leq j \leq 2n \big\}, \\
 \mfn_+ &= \text{span}_{\mathbb C}\big\{\tilde{E}_{ij}, 1 \leq i < j \leq 2n \big\}.
\end{align}
Thus our Cartan generators in the transformed basis $\tilde{e}$ are still diagonal and the raising operators are still upper triangular. As before, we introduce a set of functions
\begin{align}\label{cal-Q}
 \nu_{ij} &= \mu_{\tilde{E}_{i, 2n+1 - j}} \equiv {\cal Q}_{ji} \cr &=  \frac{1}{2} (Q_{ji}  + Q_{j, i + n} - Q_{j + n,i} - Q_{j + n, i + n})
\end{align}
for $1 \leq i,j \leq n$.  We point out that in the advanced-retarded space this is exactly the structure that appeared before in Eq.\ (\ref{e4.10}).

We now define functions $f_m$ for $1 \leq m \leq n$ in the same way as in the compact case:
\begin{align}
 f_m &= \text{Det}\big[\nu^{(m)}\big], & \nu^{(m)}&= \begin{pmatrix}
 \nu_{11} & \ldots & \nu_{1m} \\ \vdots & \ddots & \vdots \\ \nu_{m1} & \ldots & \nu_{mm} \end{pmatrix}.
\end{align}
By the same argument as for the compact case, each of the $f_m$ has the properties (\ref{e8.3}) of a highest-weight vector.

Moreover, each of the functions $f_m$ is real-valued and positive. This is seen as follows. Recalling the second expression in (\ref{mu-Y}) we have
\begin{align}
 \nu_{ij} &= \text{Tr}(g g^\dagger \Pi_{ij})
\end{align}
where
\begin{align}
 \Pi_{ij} &= \Sigma_3 \tilde{E}_{i,2n+1-j} = \tilde{E}_{2n+1-i,2n+1-j}.
\end{align}
This shows that $\nu_{ij}$ is, in fact, a single matrix element of the \emph{positive definite} matrix $gg^\dagger$ in the new basis, in the subspace spanned by $\{\tilde{e}_{2n+1-i}\}_{1 \leq i \leq n}$:
\begin{align}
 \nu_{ij} = \left( gg^\dagger \right)_{2n+1-j, 2n+1-i} .
\end{align}
Then the determinant $f_m = \text{Det}\big[\nu^{(m)}\big]$ is a principal minor of this positive definite matrix, and, therefore is positive as well. Hence, in the construction of a general highest-weight vector,
\begin{align}
\varphi_{(q_1,\ldots,q_n)} = f_1^{q_1 - q_2} f_2^{q_2 - q_3} \ldots f_{n-1}^{q_{n-1} - q_n} f_n^{q_n}
\end{align}
we may take the $q_i$ to be arbitrary complex numbers.

Notice, on the other hand, that the functions $f_m$ are the principal minors of the appropriate block of the matrix $\cal Q$, see Eq.\ (\ref{cal-Q}). Therefore, in the notation introduced in Sec. \ref{s6} for these minors, $f_m = d_m$, the highest-weight vector $\varphi_{ (q_1,\ldots,q_n)}$ is the same as the function $\varphi_{q,0}$ in Eq.\ (\ref{e6.25}).

Finally, we comment that the generalization to the supersymmetric case is straightforward. We simply need to replace all traces by supertraces and determinants by superdeterminants. Otherwise, everything goes through in the same way as before. For the purposes of this paper, it is sufficient to consider only the non-compact (boson-boson) sector of the super $\sigma$-model. In this case the powers $q_i$ can again take arbitrary complex values. We thus reproduce eigenfunctions (\ref{e6.8}) which, as explained in Sec.~\ref{s7}, are none other than the $N$-radial functions (\ref{e6.25}) of the Iwasawa-decomposition approach. If we do not restrict ourselves to the boson-boson sector, we obtain a broader class of eigenfunctions that, by the same token, will be equivalent to the plane waves (\ref{e3.9}). The powers $p_l$ corresponding to the compact sector are then non-negative integers as in Appendix~\ref{subsec:compact}.

\section{Tables of $\sigma$-model target spaces and their root systems}\label{appendix-B}

\begin{table*}
\caption{$\sigma$-model spaces. The $\sigma$-model target spaces for the localization problem fall into the large families of Riemannian symmetric superspaces. The last two columns list the compact and non-compact components of their underlying manifolds.}
\label{table:sigma-models}

\vskip 5mm
\begin{tabular}{|c|c|c|c|}
\hline
Symmetry & NL$\sigma$M
& Compact space & Non-compact space
\\
Class & (n-c$|$c) & ($ff$ sector) & ($bb$ sector)
\\
\hline
\hline
A (UE)
& AIII$|$AIII
& $\text{U}(2n)/\text{U}(n)\times\text{U}(n) \vphantom{\biggr|}$
& $\text{U}(n,n)/\text{U}(n)\times\text{U}(n) \vphantom{\biggr|}$
\\
\hline
AI (OE)
& BDI$|$CII
& $\text{Sp}(4n)/\text{Sp}(2n)\times\text{Sp}(2n) \vphantom{\biggr|}$
& $\text{O}(n,n)/\text{O}(n)\times\text{O}(n) \vphantom{\biggr|}$
\\
\hline
AII (SE)
& CII$|$BDI
& $\text{O}(2n)/\text{O}(n)\times\text{O}(n) \vphantom{\biggr|}$
& $\text{Sp}(2n,2n)/\text{Sp}(2n)\times\text{Sp}(2n) \vphantom{\biggr|}$
\\
\hline
\hline
AIII (chUE)
& A$|$A
& $\text{U}(n) \vphantom{\biggr|}$
& $\text{GL}(n,\mathbb{C})/\text{U}(n) \vphantom{\biggr|}$
\\
\hline
BDI (chOE)
& AI$|$AII
& $\text{U}(2n)/\text{Sp}(2n) \vphantom{\biggr|}$
& $\text{GL}(n,\mathbb{R})/\text{O}(n) \vphantom{\biggr|}$
\\
\hline
CII (chSE)
& AII$|$AI
& $\text{U}(n)/\text{O}(n) \vphantom{\biggr|}$
& $\begin{array}{c} \text{GL}(n,\mathbb{H})/\text{Sp}(2n) \\
\equiv \text{U}^{*}(2n)/\text{Sp}(2n) \end{array}$
\\
\hline
\hline
C (SC)
& DIII$|$CI
& $\text{Sp}(2n)/\text{U}(n) \vphantom{\biggr|}$
& $\text{SO}^*(2n)/\text{U}(n)  \vphantom{\biggr|}$
\\
\hline
CI (SC)
& D$|$C
& $\text{Sp}(2n) \vphantom{\biggr|}$
& $\text{SO}(n,\mathbb{C})/\text{SO}(n) \vphantom{\biggr|}$
\\
\hline
BD (SC)
& CI$|$DIII
& $\text{O}(2n)/\text{U}(n)  \vphantom{\biggr|}$
& $\text{Sp}(2n, \mathbb{R})/\text{U}(n) \vphantom{\biggr|}$
\\
\hline
DIII (SC)
& C$|$D
& $\text{O}(n) \vphantom{\biggr|}$
& $\text{Sp}(2n,\mathbb{C})/\text{Sp}(2n) \vphantom{\biggr|}$
\\
\hline
\end{tabular}\hfill \\
\end{table*}

\begin{table*}
\caption{
Root systems for the $\sigma$-model target spaces. We choose the system of positive roots such that, in the notation used in the table, $1\leq j<k \leq p$ and $1 \leq m < l \leq r$ (notice the opposite choice for the $bb$ and $ff$ sectors). As appropriate for our $\sigma$-model target spaces, we only consider the orthogonal groups in even dimensions. The last two columns list the coefficients of the expansions of the half-sum of positive roots $\rho = \sum_{j=1}^p c_j x_j + i \sum_{l=1}^r b_l y_l$.
}
\label{table:root-systems}
\vskip 5mm
\begin{tabular}{|c|c|c|c|c|c|c|c|c|c|c|c|c|c|}
\hline
Symmetry & NL$\sigma$M & $x_j - x_k$ & $x_j + x_k$ & $2x_j$ & $i(y_l - y_m)$ & $i(y_l + y_m)$ & $2i y_l$ & $x_j - i y_l$ & $x_j + i y_l$ & $p$ & $r$ & $c_j$ & $b_l$
\\
Class & (n-c$|$c) &  &  &  &  &  &  &  &  &  & & &
\\
\hline
\hline
A
& AIII$|$AIII
& 2 & 2 & 1 & 2 & 2 & 1 & $-2$ & $-2$ & $N$ & $N$ & $1-2j$  & $2l-1$
\\
\hline
AI
& BDI$|$CII
& 1 & 1 & 0 & 4 & 4 & 3 & $-2$ & $-2$ & $2N$ & $N$ & $-j$ & $4l-1$
\\
\hline
AII
& CII$|$BDI
& 4 & 4 & 3 & 1 & 1 & 0 & $-2$ & $-2$ & $N$ & $2N$ & $3-4j$ & $l-1$
\\
\hline
\hline
AIII
& A$|$A
& 2 & 0 & 0 & 2 & 0 & 0 & $-2$ & 0 &  $N$ & $N$ & $1-2j$  & $2l-1$
\\
\hline
BDI
& AI$|$AII
& 1 & 0 & 0 & 4 & 0 & 0 & $-2$ & 0 & $2N$ & $N$ & $\dfrac{1}{2} - j$ & $4l-2$
\\
\hline
CII
& AII$|$AI
& 4 & 0 & 0 & 1 & 0 & 0 & $-2$ & 0 & $N$ & $2N$ & $2 - 4j$ & $l - \dfrac{1}{2}$
\\
\hline
\hline
C
& DIII$|$CI
& 4 & 4 & 1 & 1 & 1 & 1 & $-2$ & $-2$ & $N$ & $2N$ & $1-4j$ & $l$
\\
\hline
CI
& D$|$C
& 2 & 2 & 0 & 2 & 2 & 2 & $-2$ & $-2$ & $N$ & $N$ & $-2j$ & $2l$
\\
\hline
BD
& CI$|$DIII
& 1 & 1 & 1 & 4 & 4 & 1 & $-2$ & $-2$ & $2N$ & $N$ & $1-j$ & $4l-3$
\\
\hline
DIII
& C$|$D
& 2 & 2 & 2 & 2 & 2 & 0 & $-2$ & $-2$ & $N$ & $N$ & $2-2j$ & $2l-2$
\\
\hline
\end{tabular}\hfill \\
\end{table*}

\pagebreak


\begin{thebibliography}{40}

\bibitem{Anderson58} P.~W.~Anderson, Phys.~Rev.~{\bf 109}, 1492 (1958).

\bibitem{AL50} {\it 50 years of Anderson localization}, ed. by E.~Abrahams (World Scientific, 2010).

\bibitem{evers08} For a review, see: F.~Evers and A.~D.~Mirlin, Rev.~Mod.~Phys.~{\bf 80}, 1355 (2008).

\bibitem{wiersma97} D.~S.~Wiersma {\it et al.}, Nature (London) {\bf 390}, 671 (1997).

\bibitem{BEC-localization} J.~Billy {\it et al.}, Nature (London) {\bf 453}, 891 (2008); G.~Roati {\it et al.}, {\it ibid.} {\bf 453}, 895 (2008).

\bibitem{faez09} S.~Faez {\it et al.}, Phys.~Rev.~Lett.~{\bf 103}, 155703 (2009).

\bibitem{lemarie10} G.~Lemari\'e {\it et al.}, Phys.~Rev.~Lett.~{\bf 105}, 090601 (2010).

\bibitem{altland97} A.~Altland and M.~R.~Zirnbauer, Phys. Rev. B {\bf 55}, 1142 (1997).

\bibitem{zirnbauer96} M.~R.~Zirnbauer, J.~Math. Phys. {\bf 37}, 4986 (1996).

\bibitem{heinzner05} P.~Heinzner, A.~Huckleberry, and M.~R.~Zirnbauer,
  Commun. Math. Phys. {\bf 257}, 725 (2005).

\bibitem{graphene-revmodphys} for a review see A.~H.~Castro Neto, F.~Guinea,
  N.~M.~R.~Peres, K.~S.~Novoselov, and A.~K.~Geim, Rev.~Mod.~Phys.~{\bf 81}, 109 (2009).

\bibitem{topins} for a review see
M.~Z.~Hasan and C.~L.~Kane, Rev.~Mod.~Phys. {\bf 82}, 3045 (2010);
 X.-L.~Qi, S.-C.~Zhang, Rev.~Mod.~Phys. {\bf 83}, 1057 (2011).

\bibitem{richardella10} A.~Richardella {\it et al.}, Science {\bf 327}, 665 (2010).

\bibitem{mirlin06} A.~D.~Mirlin {\it et al.}, Phys. Rev. Lett. {\bf 97}, 046803 (2006).

\bibitem{mirlin94}  A.~D.~Mirlin and Y.~V.~Fyodorov, Phys. Rev. Lett. {\bf 72}, 526 (1994); J. de Physique I (France) {\bf 4}, 655 (1994).

\bibitem{fyodorov04}  Y.~V.~Fyodorov and D.~V.~Savin, JETP Lett. {\bf
    80}, 725 (2004); D.~V.~Savin, Y.~V.~Fyodorov, and H.-J.~Sommers,
  {\it ibid.} {\bf 82}, 544 (2005); Y.~V.~Fyodorov, D.~V.~Savin, and
  H.-J.~Sommers, J. Phys. A: Math. Gen. {\bf 38}, 10731 (2005).

\bibitem{gruzberg11} I.~A.~Gruzberg, A.~W.~W.~Ludwig, A.~D.~Mirlin, and
  M.~R.~Zirnbauer, Phys. Rev. Lett. {\bf 107}, 086403 (2011).

\bibitem{Lee96} D-H. Lee and Z.~Wang, Phys. Rev. Lett. {\bf 76}, 4014 (1996).

\bibitem{Wang00} Z.~Wang, M.~P.~A. Fisher, S.~M.~Girvin, and J.~T.~Chalker,
  Phys. Rev. B {\bf 61}, 8326 (2000).

\bibitem{burmistrov11} I.~S.~Burmistrov, S.~Bera, F.~Evers, I.~V.~Gornyi,
  and A.~D.~Mirlin, Annals of Physics {\bf 326}, 1457 (2011).

\bibitem{Hoef86} D.~H\"of and F.~Wegner, Nucl. Phys. {\bf B275}, 561 (1986)

\bibitem{Wegner1987a} F.~Wegner, Nucl. Phys. {\bf B280}, 193 (1987).

\bibitem{Wegner1987b} F.~Wegner, Nucl. Phys. {\bf B280}, 210 (1987).

\bibitem{Wegner79} F.~Wegner, Z. Phys. B {\bf 35}, 207 (1979).

\bibitem{helgason78} S.~Helgason, {\it Differential Geometry, Lie Groups, and Symmetric Spaces} (Academic Press, 1978).

\bibitem{helgason84} S.~Helgason, {\it Groups and Geometric Analysis} (Academic Press, 1984).

\bibitem{mmz94} A.~D.~Mirlin, A.~M\"uller-Groeling, and M.~R.~Zirnbauer, Ann. Phys. (N.Y.) {\bf 236}, 325 (1994).

\bibitem{alldridge10} A.~Alldridge, Transformation Groups (2012) doi:10.1007/ s00031-012-9200-y; arXiv:1004.0732.

\bibitem{eigenfunction} The argument is as follows. Since $A$ normalizes $N$, translations by $a \in A$ transform an $N$-invariant function into another $N$-invariant function and, therefore, the $N$-radial part of an invariant differential operator is a differential operator with constant coefficients (since $A$ is abelian). Then exponential functions are clearly eigenfunctions of such operators. See an example of this in Lemma 4.1 in the Introduction in Ref. \onlinecite{helgason84}.

\bibitem{Fulton} W.~Fulton, {\it Young Tableaux: With Applications to Re\-pre\-sen\-tation Theory and Geometry}, Cambridge University Press, Cambridge, United Kingdom, 1997.

\bibitem{efetov-book} K.~B.~Efetov, {\it Supersymmetry in disorder and chaos}
  (Cambridge University Press, 1987).

\bibitem{mirlin-physrep} A.~D.~Mirlin, Phys. Rep. {\bf 326}, 259 (2000).

\bibitem{zirnbauer04} M.~R.~Zirnbauer, arXiv:math-ph/0404058.

\bibitem{janssen99} M.~Janssen, M.~Metzler, and M.~R.~Zirnbauer, Phys. Rev. B {\bf 59}, 15836 (1999).

\bibitem{klesse01} R.~Klesse and M.~R.~Zirnbauer, Phys. Rev. Lett. {\bf 86}, 2094 (2001).

\bibitem{beenakker97} C.~W.~J.~Beenakker, Rev. Mod. Phys. {\bf 69}, 731 (1997).

%\bibitem{lee95} H.~Lee, L.~S.~Levitov, A.~Yu.~Yakovets, Phys. Rev. B {\bf 51} (1995) 4079.

\bibitem{gruzberg05} I.~A.~Gruzberg, N.~Read, and S.~Vishveshwara, Phys. Rev. B {\bf 71}, 245124 (2005).

\bibitem{chalker02} J.~T.~Chalker, N.~Read, V.~Kagalovsky, B.~Horovitz,
  Y.~Avishai, and A.~W.~W.~Ludwig, Phys. Rev. B {\bf 65}, 012506 (2001).

\bibitem{bocquet00} M.~Bocquet, D.~Serban, and M.~R.~Zirnbauer, Nucl. Phys. {\bf B578}, 628 (2000); T.~Senthil and M.~P.~A.~Fisher, Phys. Rev. B {\bf 61}, 9690 (2000).

\bibitem{bera-evers-unpublished} S.~Bera and F.~Evers, unpublished.



\end{thebibliography}
\end{document}